\documentclass[12pt]{article}

\usepackage[hyphens]{url}
\usepackage{authblk} 
\usepackage[font=footnotesize]{caption} 
\usepackage{hyperref} 
\usepackage{graphicx} 
\usepackage{booktabs} 
\usepackage{multirow} 
\usepackage[font=footnotesize]{subfig} 

\title{On the Preliminary Investigation of Selfish Mining Strategy with Multiple Selfish Miners}
\date{}

\author[1]{\small Tin Leelavimolsilp}
\author[2]{Long Tran-Thanh}
\author[3]{Sebastian Stein}
\affil[ ]{Electronics and Computer Science, University of Southampton, UK}
{
  \makeatletter
  \renewcommand\AB@affilsepx{: \protect\Affilfont}
  \makeatother
  \affil[ ]{Email}
  \makeatletter
  \renewcommand\AB@affilsepx{, \protect\Affilfont}
  \makeatother
  \affil[1]{tl2f14@soton.ac.uk}
  \affil[2]{l.tran-thanh@soton.ac.uk}
  \affil[3]{ss2@ecs.soton.ac.uk}
}

\begin{document}
\maketitle

\begin{abstract}
Eyal and Sirer's selfish mining strategy has demonstrated that Bitcoin system is not secure even if 50\% of total mining power is held by altruistic miners \cite{Eyal2014}. Since then, researchers have been investigating either to improve the efficiency of selfish mining, or how to defend against it, typically in a single selfish miner setting. Yet there is no research on a selfish mining strategies concurrently used by multiple miners in the system. The effectiveness of such selfish mining strategies and their required mining power under such multiple selfish miners setting remains unknown.

In this paper, a preliminary investigation and our findings of selfish mining strategy used by multiple miners are reported. In addition, the conventional model of Bitcoin system is slightly redesigned to tackle its shortcoming: namely, a concurrency of individual mining processes. 
Although a theoretical analysis of selfish mining strategy under this setting is yet to be established, the current findings based on simulations is promising and of great interest. In particular, our work shows that a lower bound of power threshold required for selfish mining strategy decreases in proportion to a number of selfish miners. Moreover, there exist Nash equilibria where all selfish miners in the system do not change to an honest mining strategy and simultaneously earn their unfair amount of mining reward given that they equally possess sufficiently large mining power. Lastly, our new model yields a power threshold for mounting selfish mining strategy slightly greater than one from the conventional model.
\end{abstract}


\section{Introduction}

Originally invented by Nakamoto \cite{Nakamoto2008}, a blockchain is used to securely record a ledger of Bitcoin payment transactions amongst Internet users. The great success of blockchain and Bitcoin is based on an application of cryptographic puzzle, namely Proof-of-Work, and an economic incentive for miners, whom are an underlying workforce of Bitcoin system. In other words, anyone on the Internet with a sufficient amount of computational power can be a miner and solve the cryptographic puzzle to earn Bitcoin.

Due to Nakamoto's analysis, it has been widely believed that the Bitcoin system will remain secure as long as at least half of the total mining power are held by non-malicious miners, who honestly mine blocks according to the blockchain protocol \cite{Nakamoto2008}. In particular, the analysis of an attacker succeeding in double-spending his Bitcoin was modelled as a 1-dimensional random walk with an infinite time and one absorbing bound. Intuitively, the attacker will succeed if his mining power is more than half of the total power in the system.

However, a selfish mining strategy allows a malicious miner who possesses at least a third of total mining power to gain more than his fair share and could consequently disrupt Bitcoin system. As first demonstrated in Eyal and Sirer's work \cite{Eyal2014}, this strategy which secretly builds a private blockchain and strategically releases blocks causes honest miners to mine on a block that will eventually be replaced by selfish miner's. As a result, honest miners will receive less mining profit, and might stop mining or even participate in selfish mining together with the first selfish miner. In any case, a percentage of total mining power held by the selfish miner will increase and further improve the effectiveness of selfish mining strategy. Therefore, Bitcoin system is not safe against selfish mining strategy even if a half of the total mining power is held by honest miners.

Despite of such importance, there is yet no research on a selfish mining strategy simultaneously used by multiple miners. To the best of our knowledge, most works so far have only focused on one malicious miner using selfish mining strategy and the others employing honest mining strategy \cite{Eyal2014,Gobel2016,Sapirshtein2017,Nayak2016,Zhang2017}. However, a number of miners in Bitcoin system could use the selfish mining strategy at the same time. Whether the selfish mining strategy is still effective in such situation has not yet been investigated. 

In addition, none of previous works so far has considered a concurrency resulted from individually mining and could misestimate a power threshold of selfish mining strategy as a result. In other words, a mining process in Bitcoin system has been widely modelled as a single entity randomly assigning new blocks to their owners according to their mining powers \cite{Eyal2014,Zhang2017}. On the contrary, miners in practical individually mine on their locally-stored blockchain and consequently create new blocks independently. Therefore, the widely used model might not well represent Bitcoin system and thus previous findings of the power threshold might be lower or greater than the actual one. 

For these reasons, we have started a preliminary investigation into an effectiveness of selfish mining strategy employed by multiple miners with our new model of Bitcoin system. In essence, our model better reflects the concurrency of individually mining and further assumes that there is always at least an honest miner in the system. Such honest miner collectively represents altruistic miners in Bitcoin community who believe in a long-term economic impact of a security attack and consequently adhere to the original blockchain protocol \cite{Buterin2013}.

Despite a lack of theoretical analysis and some limitations, our findings is useful and provides an insight toward an effectiveness of selfish mining strategy used by multiple miners. We also consider our work to be a complement to Kiayias's game-theoretical analysis of blockchain mining \cite{Kiayias2016}. In particular, our work provides a number of contributions as follows.

\begin{enumerate}
 \item Under a system where there are multiple selfish miners, the least amount of mining power that could allow a selfish miner to earn his unfair amount of mining reward becomes lesser in proportion to a number of selfish miners in the system.
 \item Given a specific power configuration or a specific allocation of mining power, there exist Nash equilibria where multiple miners use a selfish mining strategy and simultaneously gain their mining reward greater than their percentage of total mining power.
 \item From a perspective of security, an amount of mining power required to completely secure against a selfish mining strategy remains the same regardless of a number of selfish miners in the system.
 \item Due to a simplified model of Bitcoin system that are widely used in literatures, all estimated mining power to effectively mount a selfish mining strategy and other attacks so far could be slightly underestimated.
\end{enumerate}

In the rest of this paper, our work are presented as follows. Firstly a literature related to our study is reviewed. Then we briefly describe our model and specifically point out any difference to the conventional model of Bitcoin system. Subsequently our simulation settings and their results are demonstrated and discussed. Finally we concludes this paper with our findings and future work.

\section{Related Work}

After Eyal and Sirer's seminal work \cite{Eyal2014}, a number of studies have advanced the research of selfish mining strategy. For example, G\"{o}bel and his colleagues extended the original study by using a spatial Poisson point process to model a network of miners to better account for an effect of network delay \cite{Gobel2016}. Their findings showed a relatively high number of blocks in a blockchain and consequently a great amount of mining reward for every miner if there is no miner the using selfish mining strategy. Nonetheless, the selfish mining strategy still allows a miner to gain his mining reward greater than his percentage of total mining power once his mining power is sufficiently large.

On the one hand, some studies have improved the original selfish mining strategy to be more effective and consequently gain higher amount of mining reward than the original one \cite{Sapirshtein2017,Nayak2016}. In addition to optimising the original selfish mining strategy, a combination with other attacks such as an eclipse attack could be used to further increase its success rate \cite{Nayak2016}. As a result, a safety level of mining power required for Bitcoin system to be secure against the selfish mining strategy became lower than $1/3$ of the total mining power.

On the other hand, a number of researches proposed various methods that improve blockchain protocol to better resist the selfish mining strategy, though they were difficult to carry out in practical \cite{Zhang2017}. In particular, these methods increased a power threshold or an amount of mining power required for successfully mounting selfish mining strategy to some degree. However, their methods including Zhang and Preneel's required a good coordination amongst the majority of miners to adopt them at a specific point of time to prevent undesirable forking of the blockchain. Consequently, they were hard to implement in Bitcoin system.

Whilst a game-theoretical analysis could provide a great insight into miner's decision whether to mount a selfish mining strategy, game theory is based on an assumption of strong rationality and therefore might not reflect altruistic miners in Bitcoin system. With regard to mining strategies that strategically publish their hidden blocks, Kiayias and his colleagues demonstrated that every miner will use an honest mining strategy if no one possesses mining power greater than 30.8\% of the total mining power \cite{Kiayias2016}. However, their model was based on an assumption of rational miners; in other words, miners always change their mining strategy to the most beneficial one. Such assumption did not reflect altruistic miners, who adhere to the original mining software due to a fear of long-term economic impact resulted from any security attack.

To the best of our knowledge, there is no research that considers multiple miners simultaneously employing a selfish mining strategy. In practical, a selfish miner might not be able to immediately switch back and forth between an honest mining strategy and the selfish mining strategy to best maximise his mining reward. In other words, they might continuously use the selfish mining strategy for some periods even if it is less profitable than the honest mining strategy. A possible cause of selfish miner's irrational decision could be a lack of information; e.g. a current distribution of mining powers which could be heavily fluctuating.

Furthermore, all works so far have modelled Bitcoin system in a slightly unrealistic manner and could consequently misestimate a power threshold required for the selfish mining strategy. That is, their models assumed that there is always one miner successfully generating a block at each period. However, miners in practical individually and concurrently mine blocks on their locally-stored blockchain. As a result, there could be more than one block generated in each period. Therefore, the actual power threshold might be greater or lower than the power threshold that has been known so far.

\section{Our Model of Bitcoin System}

To tackle the shortcomings previously mentioned, the widely used model of Bitcoin system is redesigned to reflect miners concurrently mining on their locally-stored blockchains. In addition, an honest miner who collectively represents altruistic miners in Bitcoin system is included into our model. As a consequence, our model better represents Bitcoin system. In particular, our model is described as follows.

\begin{itemize}
 \item Each miner is denoted by $i \in \left\lbrace 1,2,...,N \right\rbrace $, in which every miner $i \in \left\lbrace 1,2,...,N-1 \right\rbrace $ always uses the original selfish mining strategy \cite{Eyal2014} and the $N$-th miner uses an honest mining strategy \cite{Nakamoto2008}. For the sake of simplicity, all altruistic miners are treated as one honest miner and we focus on the effect of varying number of selfish miners and their mining powers.
 \item For each miner, his mining process on locally-stored blockchain is a Bernoulli trial with a probability of successfully creating a block $ p_i = m_i \times d $, where: 
 \begin{itemize}
  \item $ m_i \in \left[ 0,1 \right] $ denotes \textit{a relative mining power} or a proportion of total mining power that miner $i$ possesses, and we refer to an allocation of mining power to every miner as \textit{a power configuration} $ M = \left\lbrace m_i | i \in \left\lbrace 1,2,...,N \right\rbrace \right\rbrace $ in which $ \sum_{i=1}^{N} m_i = 1 $;
  \item $ d \in \left[ 0,1 \right] $ represents a difficulty level of creating blocks, i.e., a proportion of nonces that yield a hash value lower than a target value given that all miners create the same block. Analogously, a high value of $d$ represents Bitcoin system with a high target value and vice versa for a low value of $d$.
 \end{itemize}
 \item Utility function $ U_i $ of each miner is defined as \textit{a relative reward} or a proportion of his blocks in the longest blockchain and thus $ U_i \in \left[ 0 , 1 \right] $ . 
 \item Every miner is assumed to be directly connected to all other miners and there is no communication delay in the network. Consequently, any broadcast message regarding a discovery of new blocks is instantaneously received. However, if there is a number of messages simultaneously sent to the same recipient, an ordering of messages that will be sequentially received is randomly chosen and thus all messages have an equal probability of being first received.
\end{itemize}

Note that our model neither regards selfish miner's network capability as Eyal and Sirer's model \cite{Eyal2014} nor favours altruistic miner's one. Since network capability and communication delay could greatly affect the effectiveness of selfish mining strategy, a study involving them should be systematically carried out and thus it is considered as a next step of this work.

Finally, it should be pointed out that a major difference between our model and the conventional model is a mining process. In particular, the mining process in the conventional model was modelled as a global entity which generates new blocks and randomly assign them to miners according to their mining powers. On the other hand, our model mimics all of individual mining processes and allows a number of miners to simultaneously generate their blocks. For further details of the conventional model, readers should refer to Eyal and Sirer's work \cite{Eyal2014}.

\section{Simulation and Results}

A series of discrete event simulations were carried out to observe an effectiveness of selfish mining in three different settings: namely one selfish miner, two selfish miners, and three selfish miners. In particular, a relative mining power of selfish miner is varied by 0.01. Furthermore, there is one honest miner who has the rest of mining power left from selfish miners'. Each simulation which runs for 200,000 timesteps is then repeated 100 times to calculate an average relative reward $ E \left[ U_i \right] $ and its 95\% confidence interval. 

It is worth mentioning that parameter $d$ of our model is fixed to 0.5 in all simulations. Since $d = 0.5$ allows blocks to be frequently generated, the resulted blockchain will be sufficiently long to extract meaningful results. Together with 200,000 timesteps, a length of the resulted blockchain varies from 40,000 - 90,000 blocks, which is analogous to 11 - 13 months' worth of Bitcoin's blockchain.

In the following, simulation results will be shown and described for each setting. Except for the case of three selfish miners, we compare simulation results of our model and the conventional model's. Furthermore, a simulation result of our model when there is no selfish miner is included to demonstrate a baseline behaviour of our model.

\subsection{No Selfish Miner}

\begin{figure}
 \centering
 \includegraphics[width=0.65\textwidth]{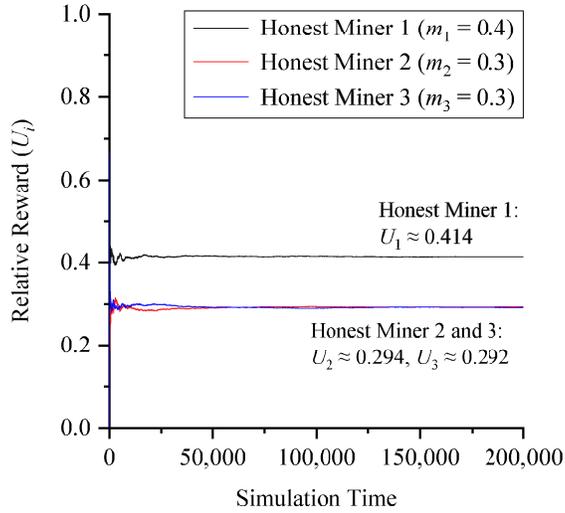}
 \caption{A line plot demonstrating a convergence of miners' reward in our model without any selfish miner.}
 \label{fig:allhm}
\end{figure}

In contrast to the conventional model, our model shows an interesting aspect of the actual mining process: that is, a miner with the greatest amount of mining power was given a mining reward slightly larger than his relative power and vice versa for the other miners. As shown in figure \ref{fig:allhm}, an honest miner with a relative mining power of 0.4 received an amount of relative reward 0.414, whereas all other miners (each with a power of 0.3) got slightly lesser than their relative powers.

Nevertheless, our model captures one of the main characteristics of Bitcoin mining. That is, a miner receives his mining reward in proportion to his mining power.

\subsection{One Selfish Miner}

\begin{figure}
 \centering
 \includegraphics[width=0.65\textwidth]{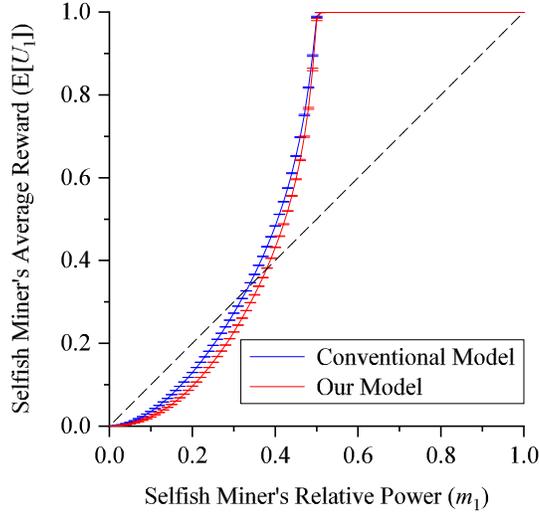}
 \caption{A line plot comparing selfish miner's average reward and its 95\% confidence interval between the conventional model and our model in one-selfish-miner setting. A black dashed line indicates an amount of reward the selfish miner would get if he had followed an honest mining strategy.}
 \label{fig:1sm}
\end{figure}

As demonstrated in figure \ref{fig:1sm}, a selfish miner who has mining power exceeding a threshold in both models gains a relative reward more than his relative power, however our model yields the threshold slightly higher than the conventional model's. In particular, a selfish miner requires a relative power of at least 0.34 to effectively use selfish mining strategy in the conventional model, whereas a power of at least 0.38 is needed in our model.

Such difference could be due to our model's concurrent mining processes, which further results in a greater demand of mining power for a selfish mining strategy to successfully create the longest blockchain.

\subsection{Two Selfish Miners}

\begin{figure}
 \centering
 \subfloat[]{
  \centering
  \includegraphics[width=0.475\textwidth]{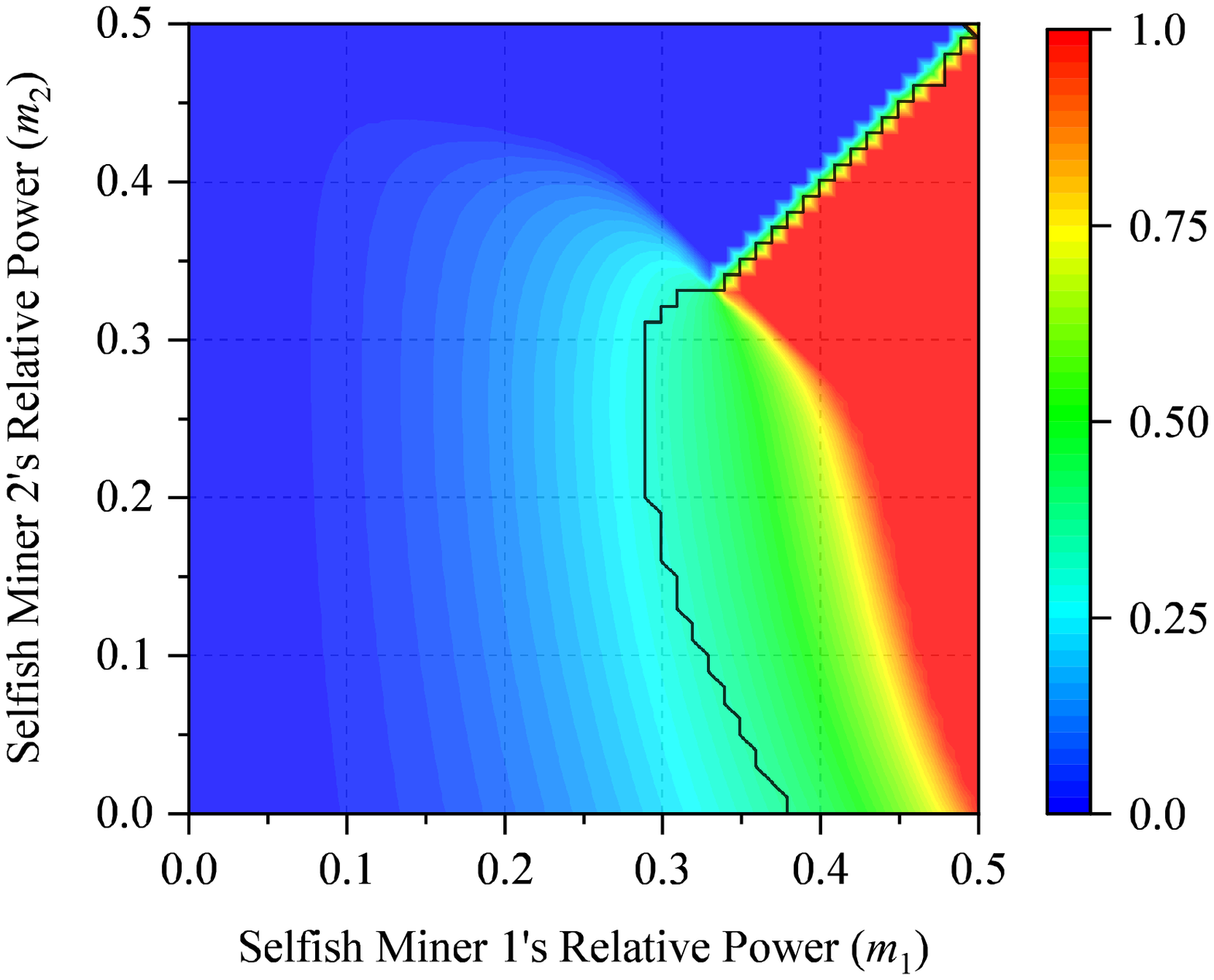}
  \label{fig:2sm_reward}
 } 
 \subfloat[]{
  \centering
  \includegraphics[width=0.475\textwidth]{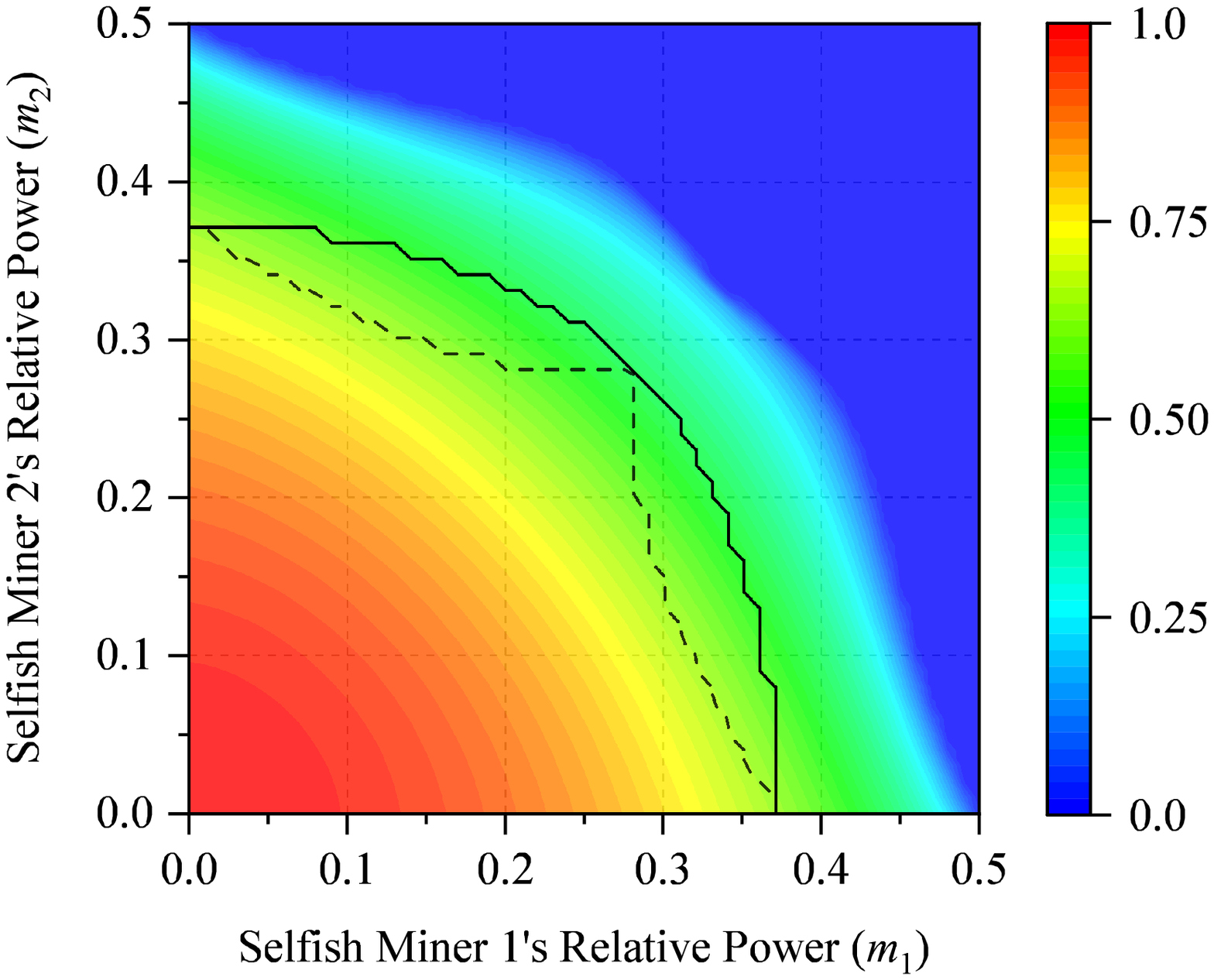}
  \label{fig:2hm_reward}
 } \\
 \subfloat[]{
  \centering
  \includegraphics[width=0.475\textwidth]{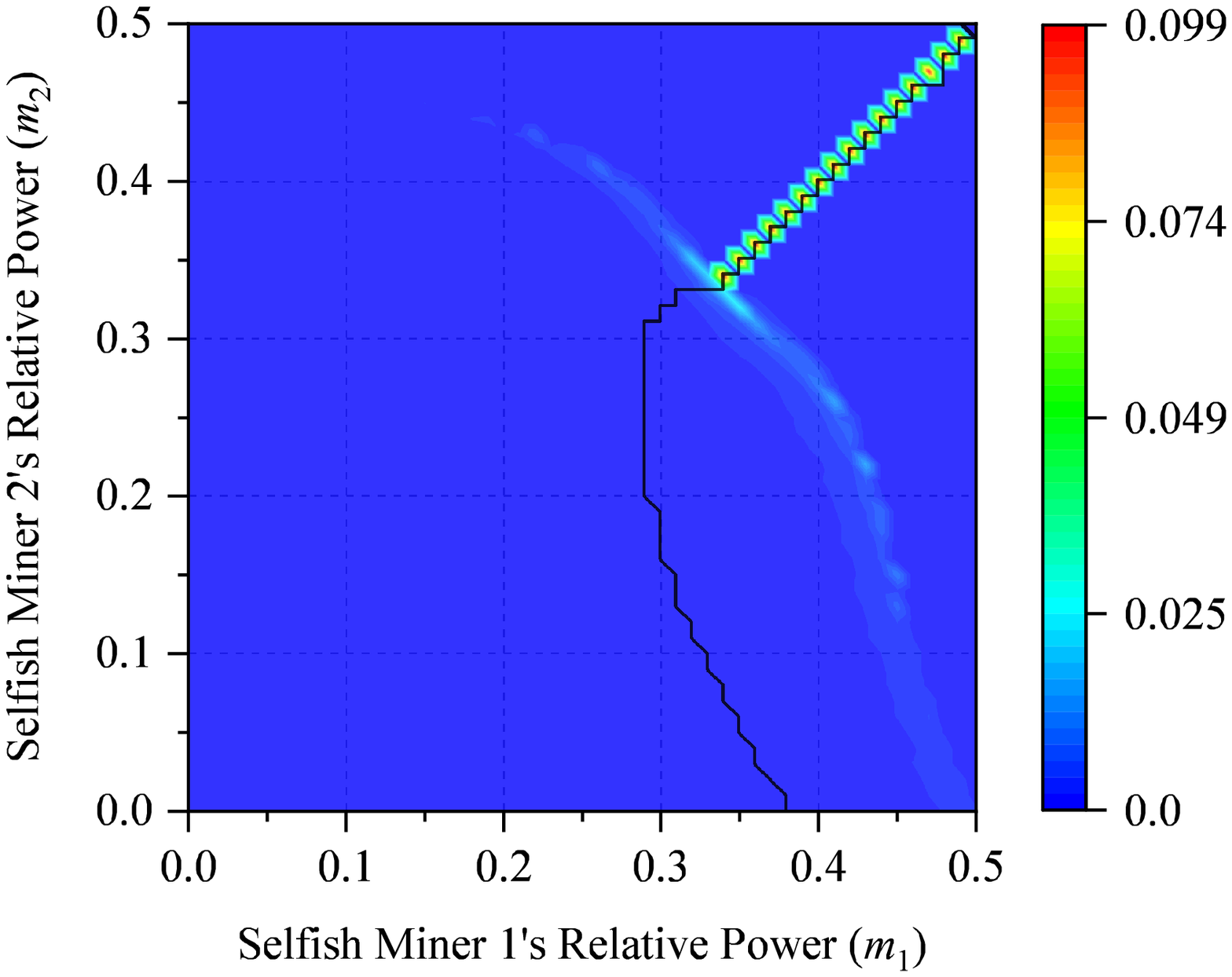}
  \label{fig:2sm_ci}
 }
 \subfloat[]{
  \centering
  \includegraphics[width=0.475\textwidth]{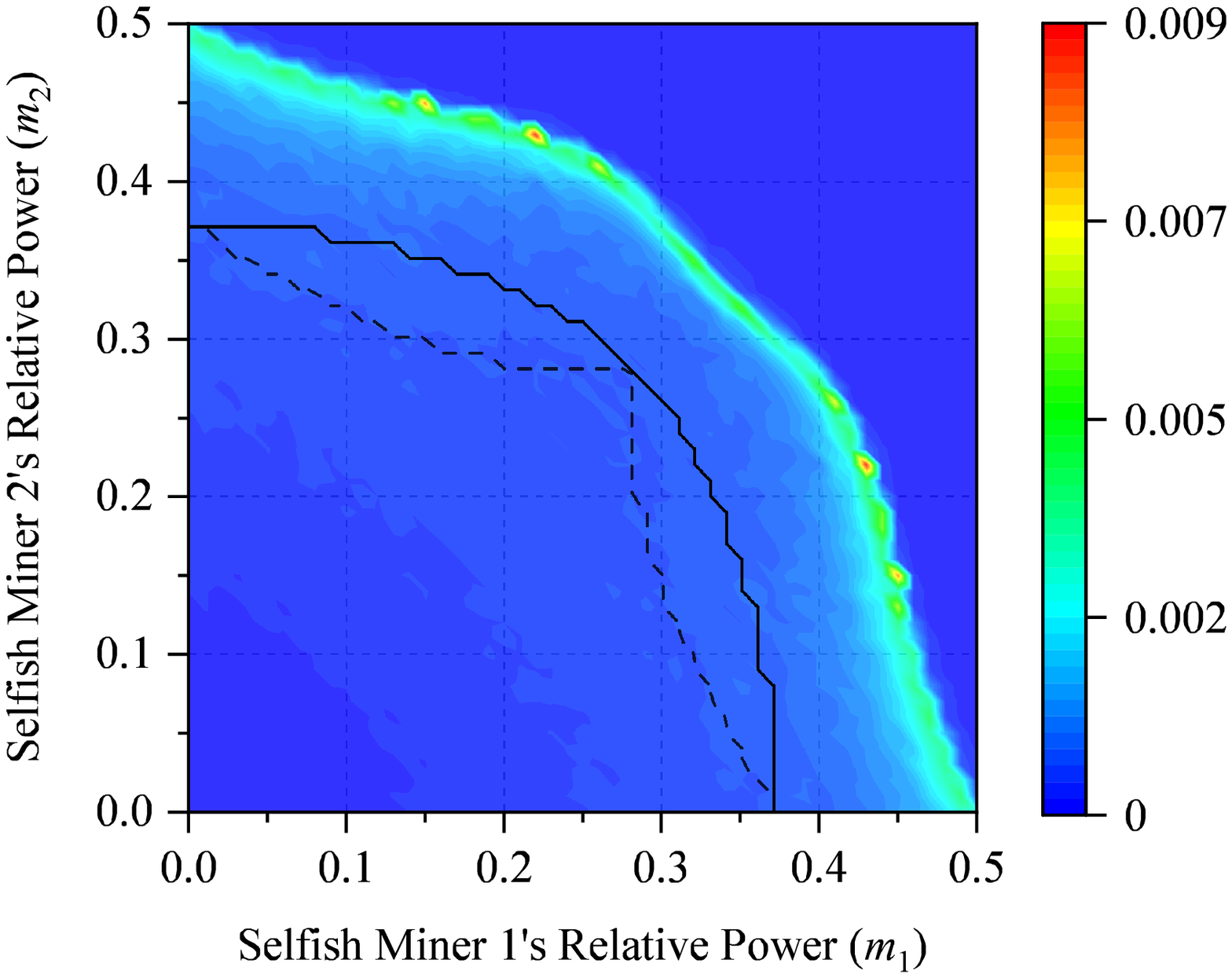}
  \label{fig:2hm_ci}
 }
 \caption{Heat maps of selfish miner 1's average reward (a), honest miner's average reward (b), and their 95\% confidence intervals respectively (c,d) in our model with two selfish miners. In each plot of selfish miner 1's (a,c), a black line separates power configurations that yield mining reward greater than his power to the right side and vice versa to the left side of the plot. Similarly, a black solid line in each plot of honest miner's (b,d) separates power configurations that yield reward at least equal to his relative power to the lower left side and vice versa to the upper right side of the plot. Also, a dashed line separates power configurations where there is at least 1 selfish miner earning his unfair amount of reward to the upper right side and vice versa to the lower left side of the plot. Note that an intersected area between these two lines indicates power configurations where the honest miner receives his fair share of reward and at least one selfish miner earns his extra amount of reward as well.}
 \label{fig:2_reward}
\end{figure}

In contrast to the previous setting, a power threshold of selfish mining strategy under this setting is comparatively low. As shown in figure \ref{fig:2sm_reward}, a power configuration that has the least amount of selfish miner 1's power yet still allows him to gain his unfair amount of reward is $ M = \left\lbrace 0.29, m_2, m_3 \right\rbrace $ where $ m_2 \in \left[ 0.2, 0.31 \right] $ and $ m_3 $ has the rest. On the other hand, other power configurations $ M' = \left\lbrace 0.29, m_2, m_3 \right\rbrace $ where $ m_2 \notin \left[ 0.2, 0.31 \right] $ do not yield extra reward for selfish miner 1. A plausible cause for the latter might be an insufficient amount of selfish miner's power to frequently win against either an honest miner or the another selfish miner in a race to create the longest blockchain.

\begin{figure}
 \centering
 \subfloat[]{
  \centering
  \includegraphics[width=0.5\textwidth]{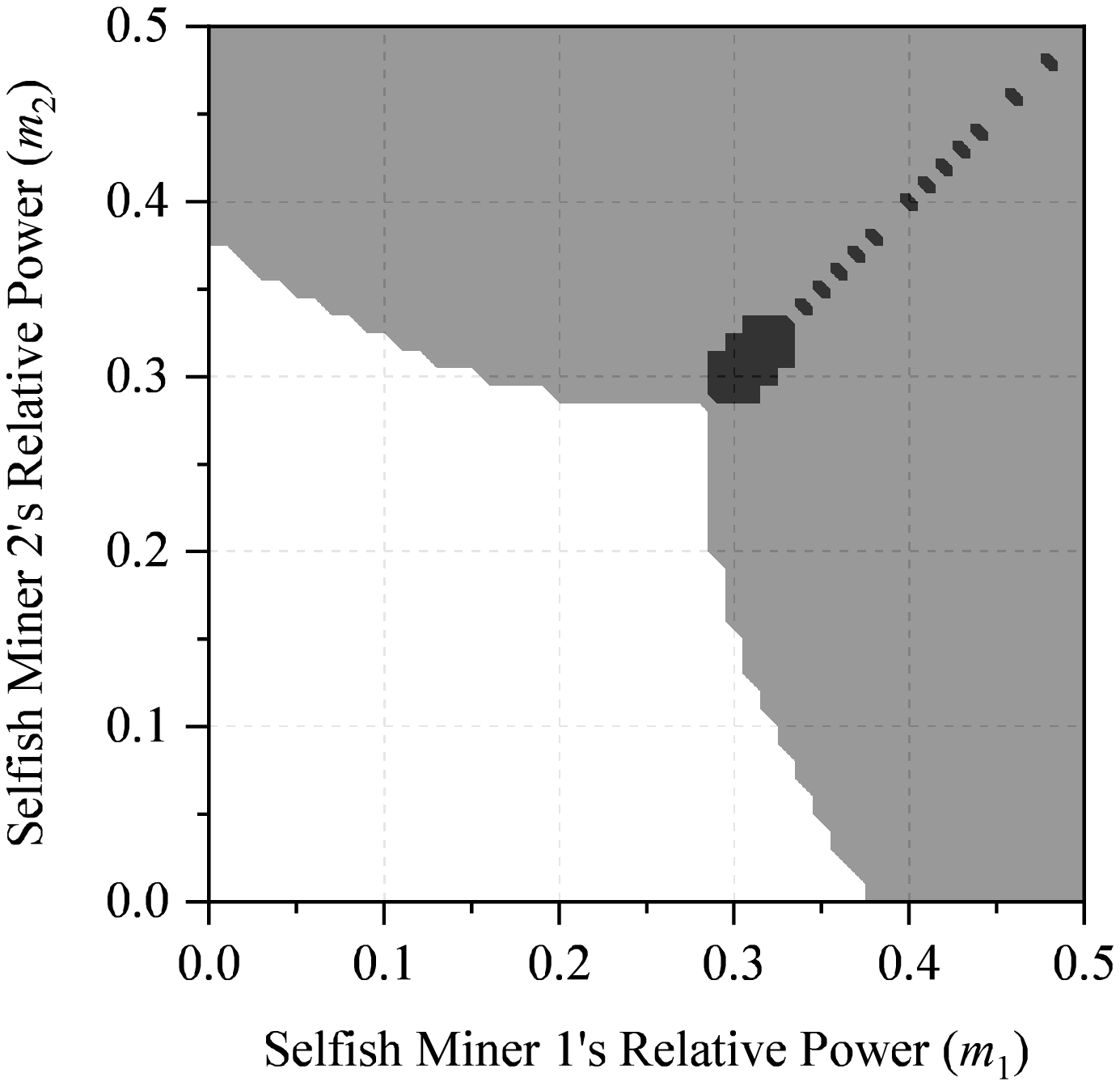}
  \label{fig:2sm_effective_geoDist}
 } 
 \subfloat[]{
  \centering
  \includegraphics[width=0.5\textwidth]{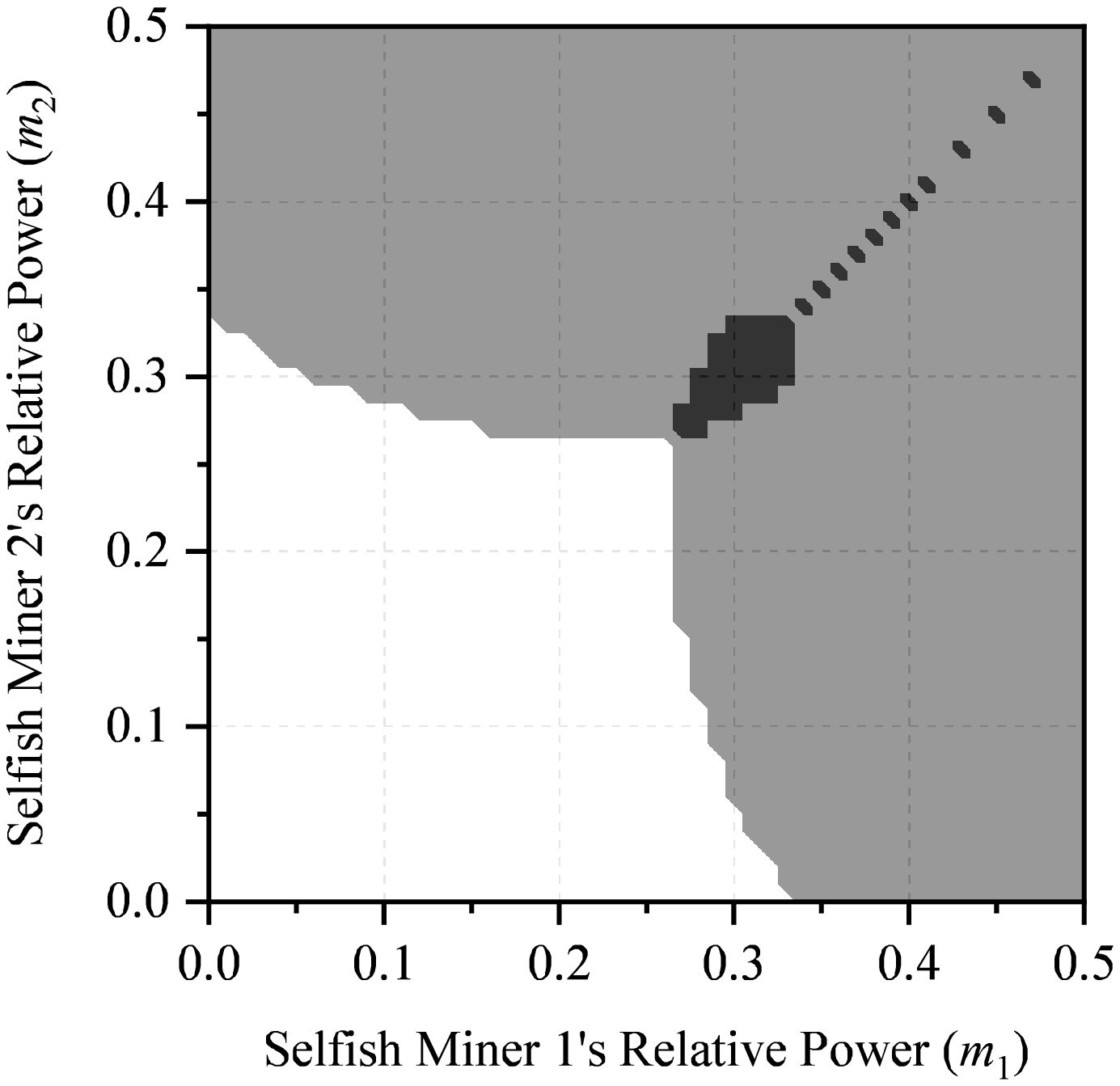}
  \label{fig:2sm_effective_bernDist}
 }
 \caption{Contour plots comparing power configurations that a selfish mining strategy is effective between our model (a) and the conventional model (b) with 2 selfish miners. A gray area indicates configurations that only 1 selfish miner gains his unfair amount of mining reward, whereas a black area indicates configurations that both selfish miners simultaneously earn their extra amounts of mining reward.}
 \label{fig:2sm_effective_config}
\end{figure}

\begin{figure}
 \centering
 \subfloat[]{
  \centering
  \includegraphics[width=0.5\textwidth]{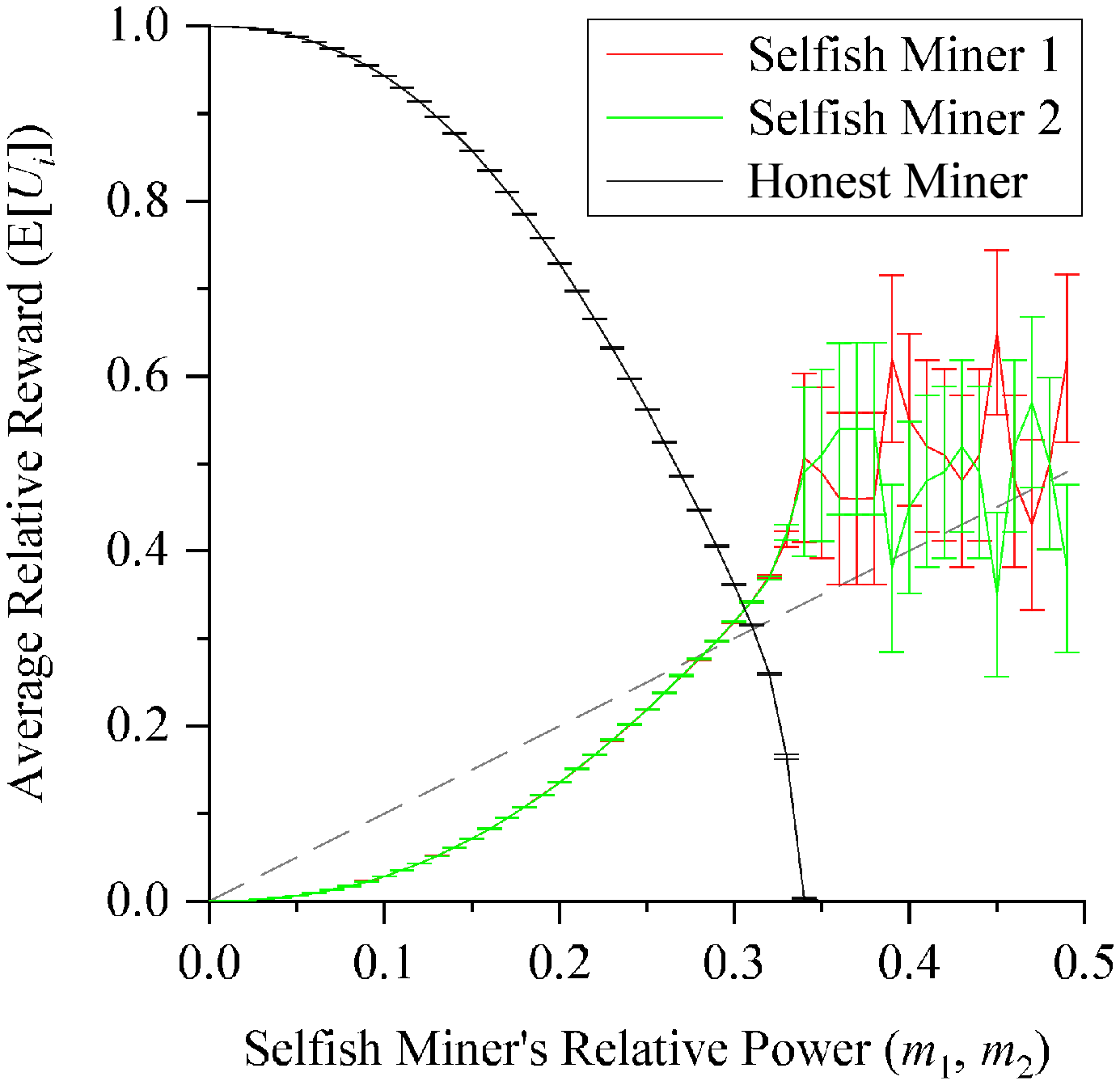}
  \label{fig:2sm_symm_100rep}
 } 
 \subfloat[]{
  \centering
  \includegraphics[width=0.5\textwidth]{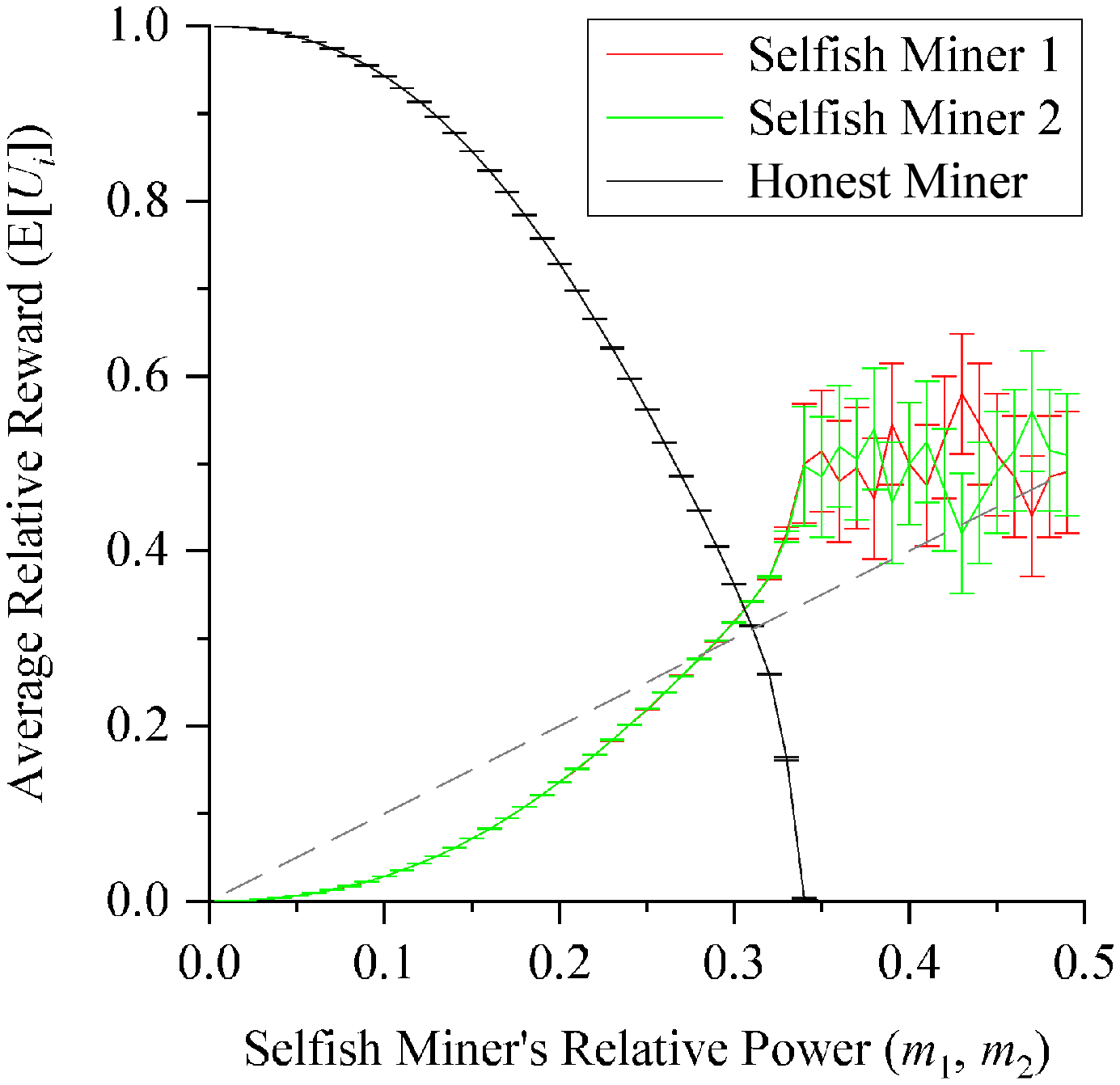}
  \label{fig:2sm_symm_200rep}
 } \\
 \subfloat[]{
  \centering
  \includegraphics[width=0.5\textwidth]{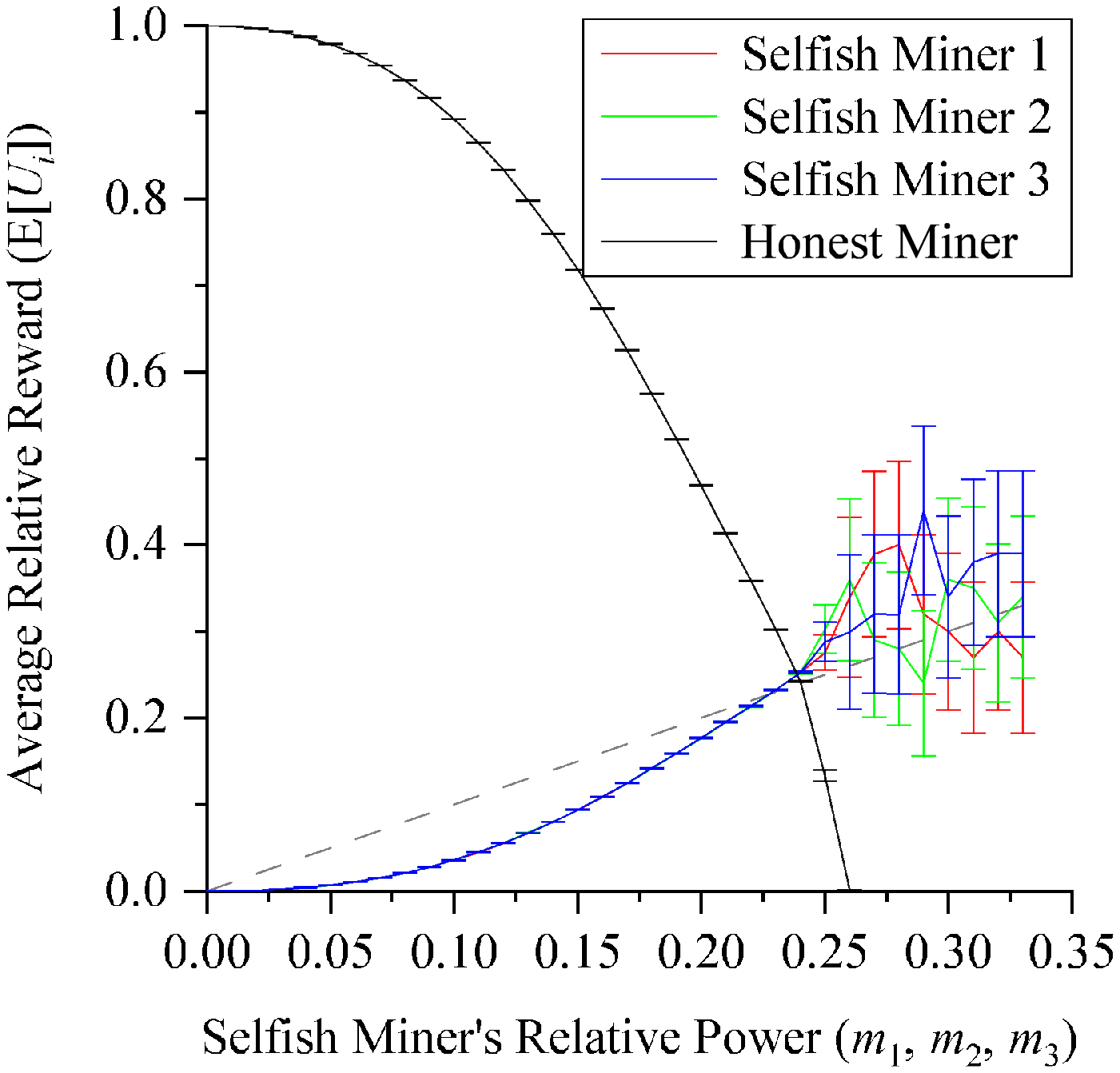}
  \label{fig:3sm_symm_100rep}
 } 
 \subfloat[]{
  \centering
  \includegraphics[width=0.5\textwidth]{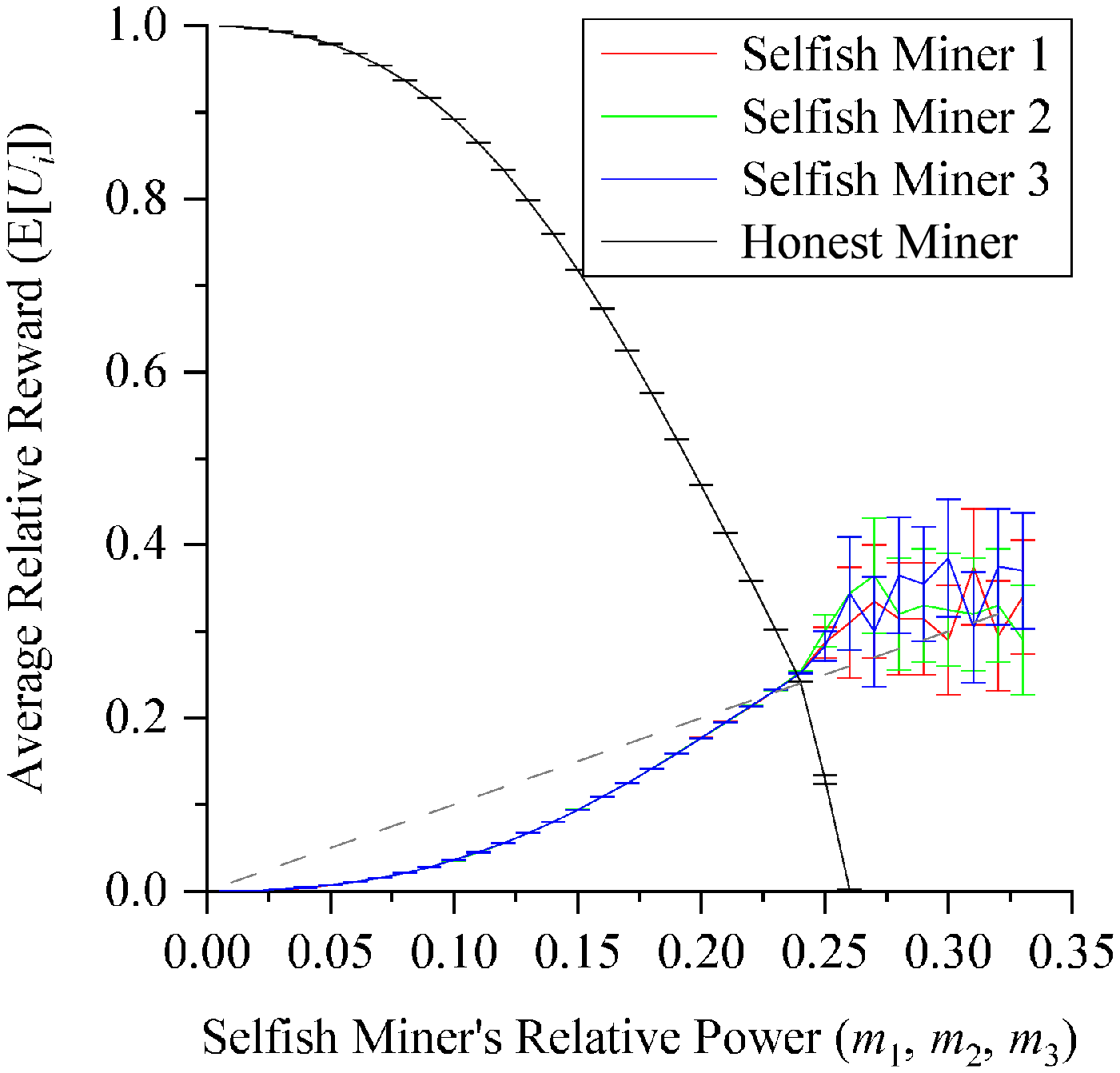}
  \label{fig:3sm_symm_200rep}
 }
 \caption{Line plots demonstrating miners' average reward and their 95\% confidence intervals of a power configuration where all selfish miners possess an equal amount of mining power. All upper plots (a,b) are simulation results of our model with 2 selfish miners, whilst all lower plots (c,d) are results of our model with 3 selfish miners. All left plots (a,c) are simulation results with 100 repetitions each, whereas all right plots (b,d) are results with 200 repetitions each.}
 \label{fig:sm_symm}
\end{figure}

Surprisingly, both selfish miners can simultaneously gain their extra rewards in some specific power configurations. As demonstrated in figure \ref{fig:2sm_effective_geoDist} and \ref{fig:2sm_symm_200rep}, a power configuration $ M = \left\lbrace m_1, m_2, m_3 \right\rbrace $ where $ m_1 \in \left[ 0.29, 0.49 \right] $ and $ m_1 = m_2 $ allows both selfish miners to earn their relative rewards higher than their relative powers. However, their reward amounts become unstable for any power configuration where both selfish miners' power equally exceeds 0.33. Specifically, their reward amounts could be lesser than their relative powers if they equally possess relative power greater than or equal to 0.41.

It is also worth mentioning that (i) a system under this setting is completely vulnerable to selfish mining strategy for any power configuration where an honest miner has relative mining power less than 0.44, (ii) as portrayed in figure \ref{fig:2hm_reward}, there are some power configurations $ M = \left\lbrace m_1, m_2, m_3 \right\rbrace$, $ m_3 \in \left[ 0.44, 0.61 \right] $ that allow an honest miner to retain his reward in proportion to his power whilst also let one selfish miner earn his unfair amount of reward, and (iii) a power threshold to effectively use selfish mining strategy under the conventional model is slightly underestimated in comparison to our model's. A simulation result of the latter is demonstrated in appendix \ref{appA}.

\subsection{Three Selfish Miners}

\begin{figure}
 \centering
 \subfloat[]{
  \centering
  \includegraphics[width=0.45\textwidth]{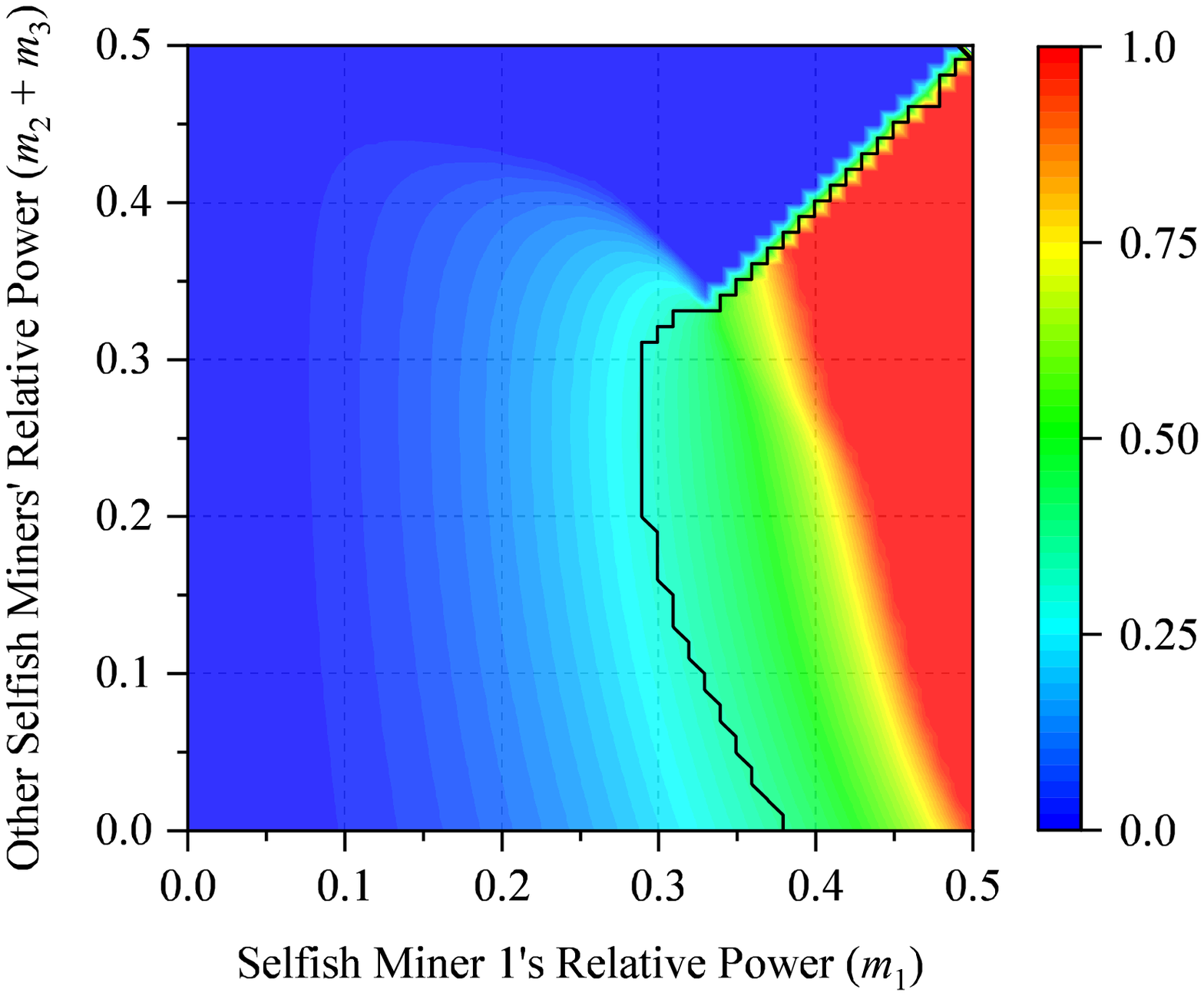}
  \label{fig:3sm_min}
 } 
 \subfloat[]{
  \centering
  \includegraphics[width=0.45\textwidth]{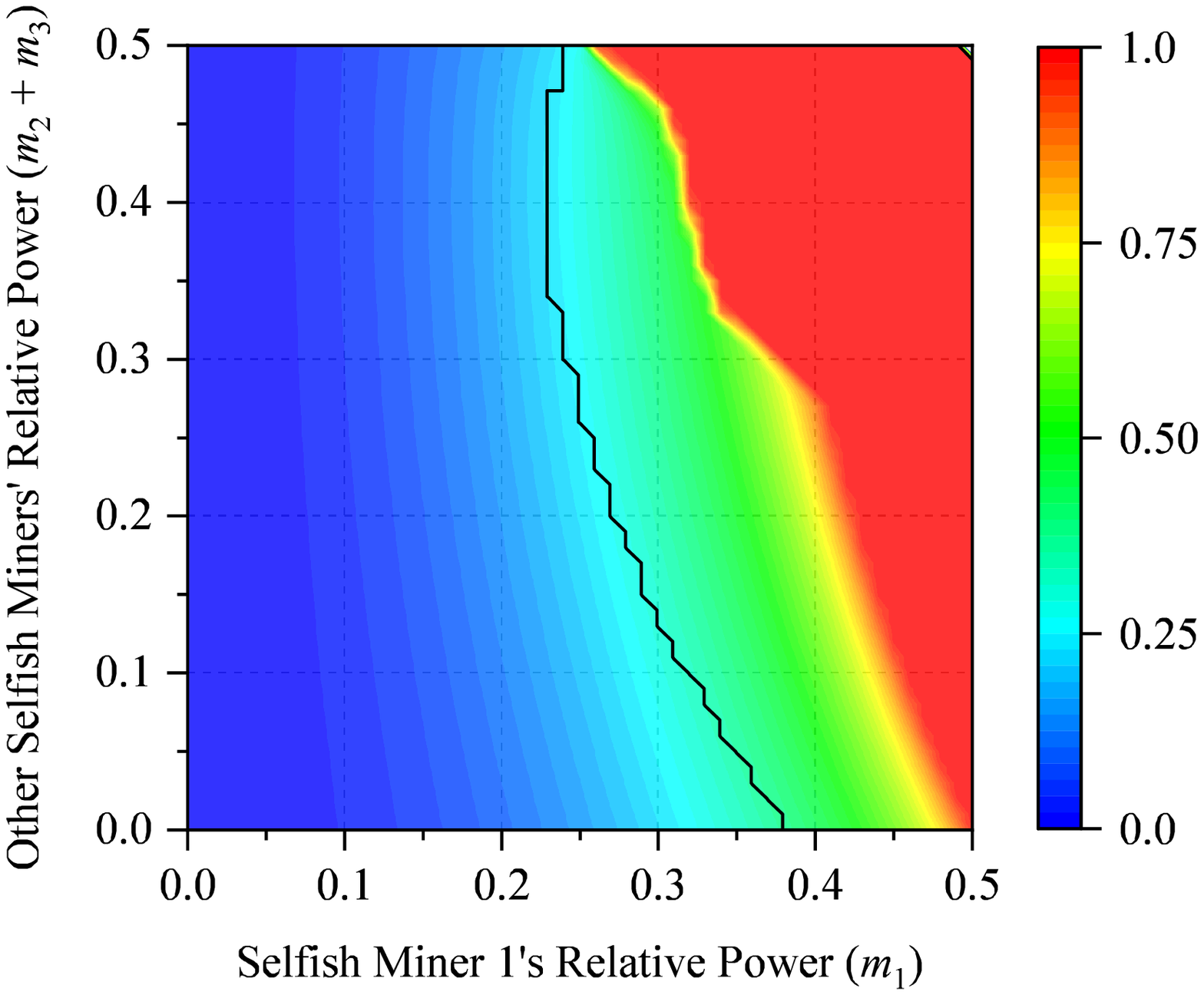}
  \label{fig:3sm_max}
 } \\
 \subfloat[]{
  \centering
  \includegraphics[width=0.45\textwidth]{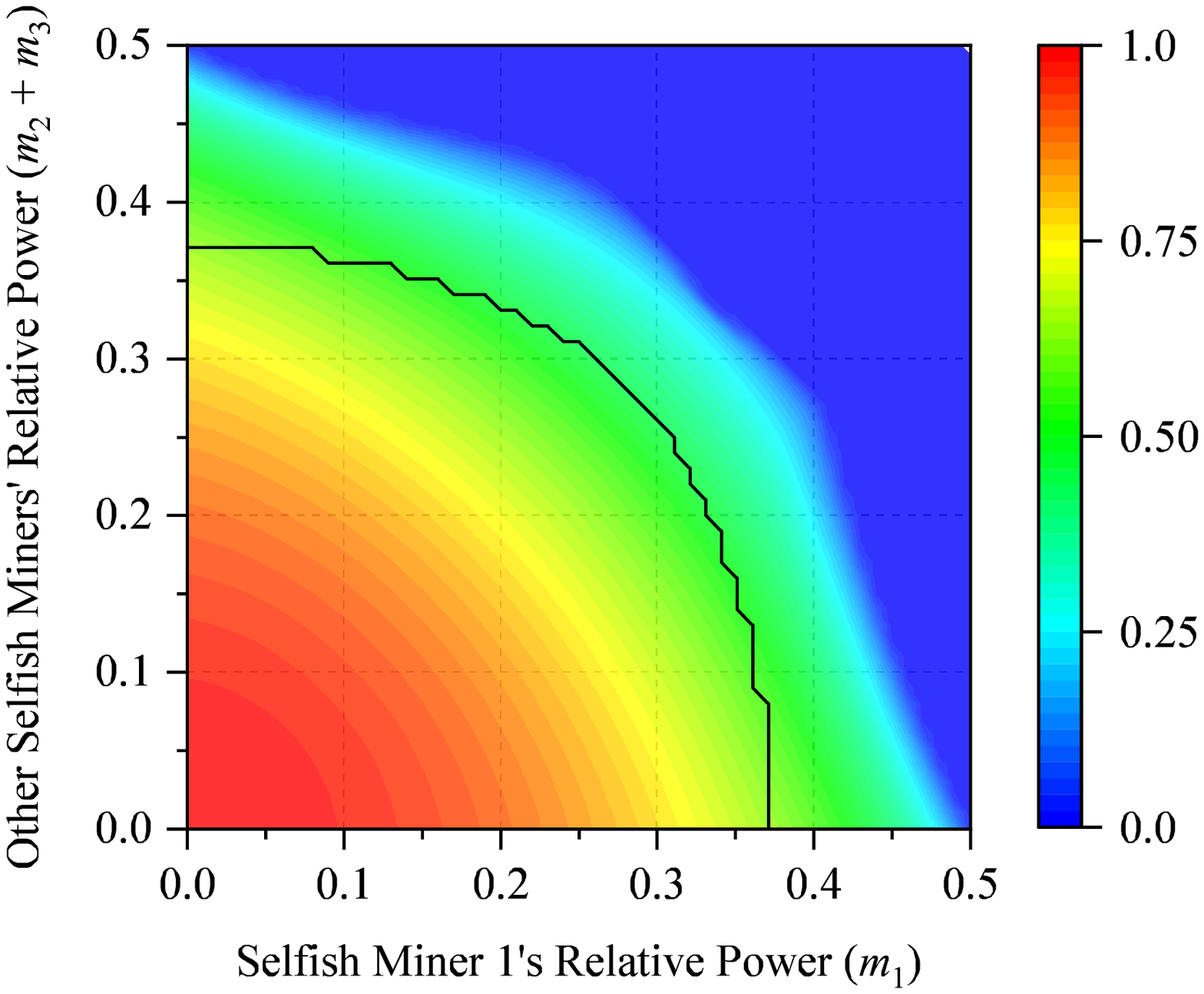}
  \label{fig:3hm_min}
 } 
 \subfloat[]{
  \centering
  \includegraphics[width=0.45\textwidth]{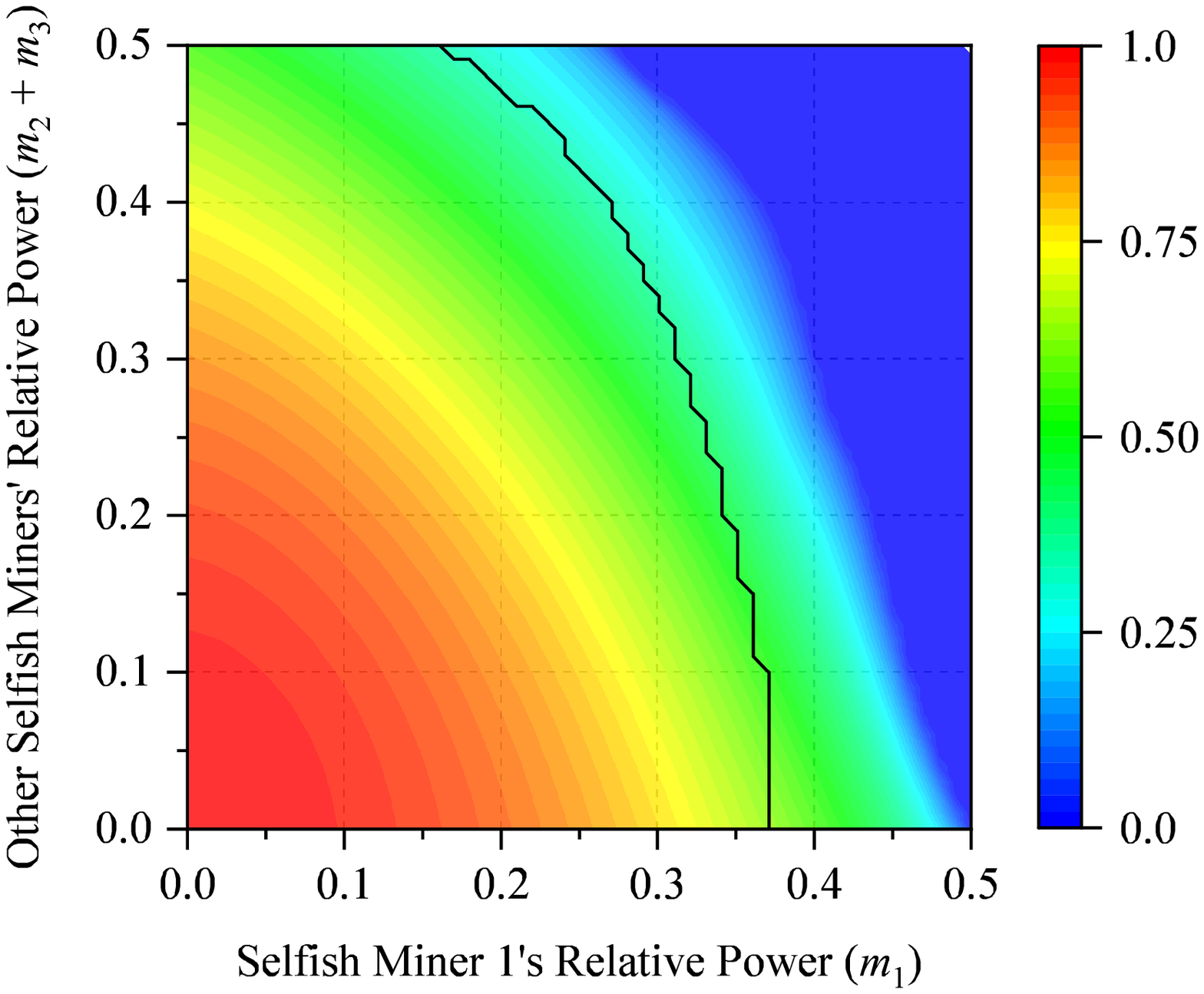}
  \label{fig:3hm_max}
 } \\
 \subfloat[]{
  \centering
  \includegraphics[width=0.23\textwidth]{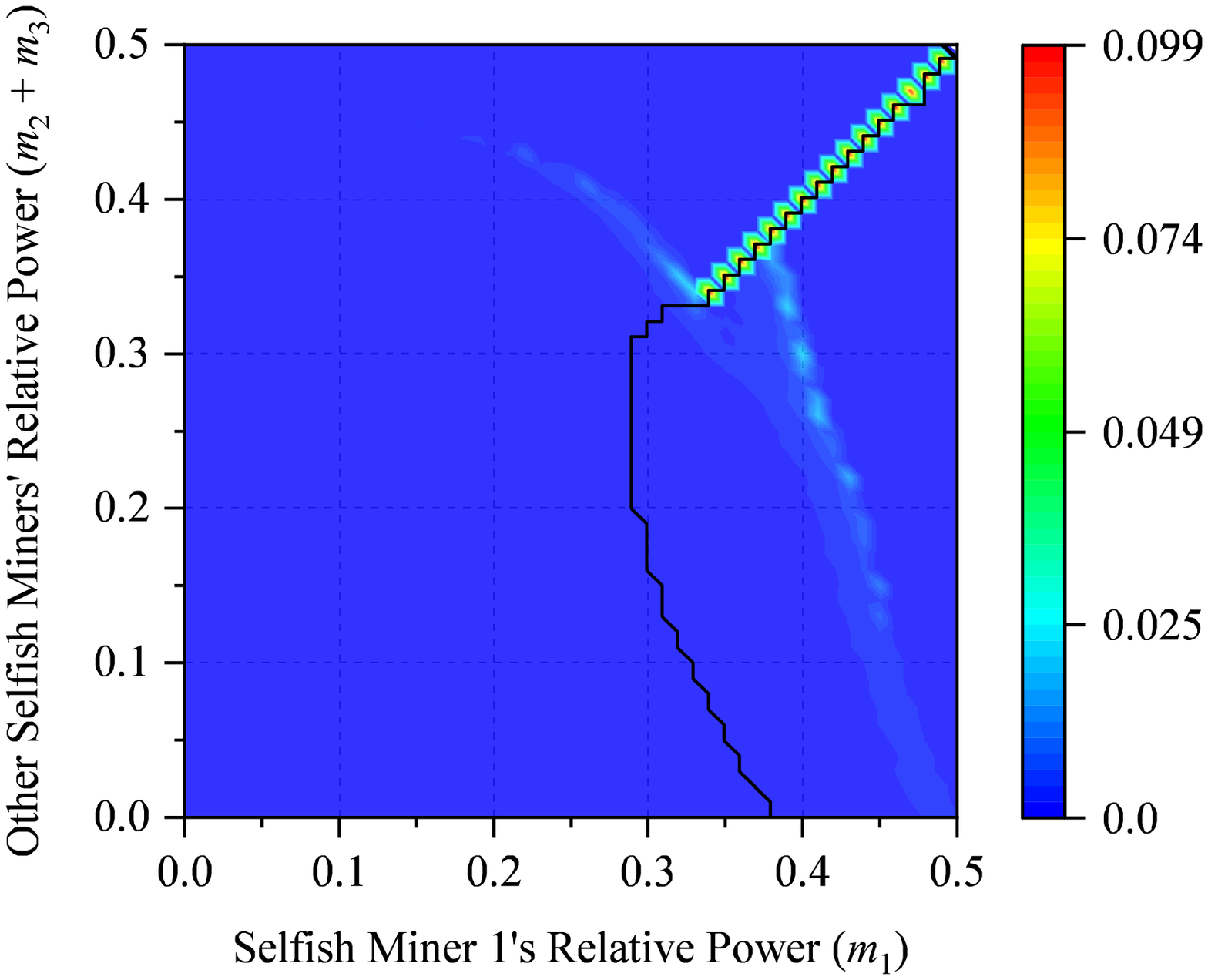}
  \label{fig:3sm_minCI}
 } 
 \subfloat[]{
  \centering
  \includegraphics[width=0.23\textwidth]{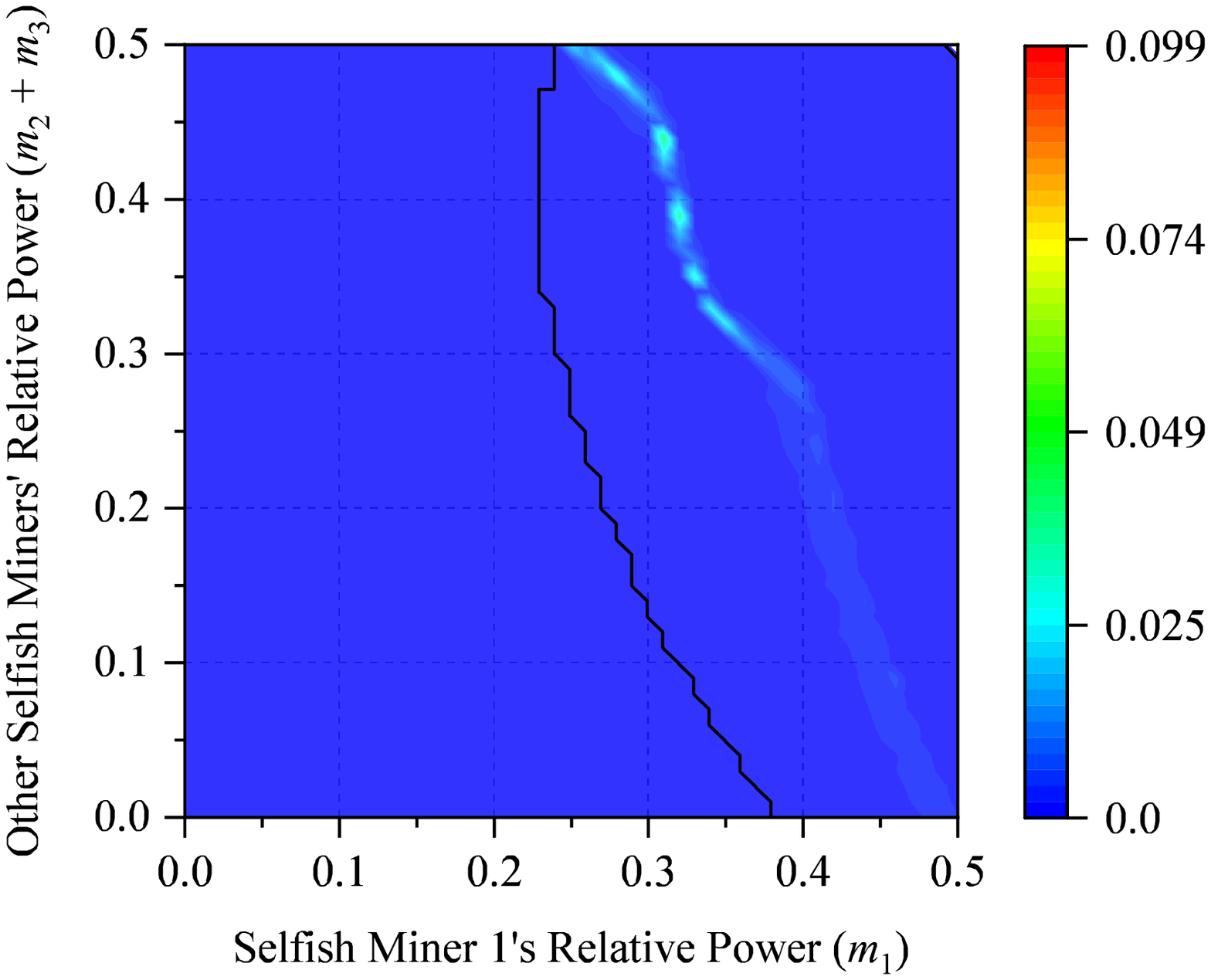}
  \label{fig:3sm_maxCI}
 }
 \subfloat[]{
  \centering
  \includegraphics[width=0.23\textwidth]{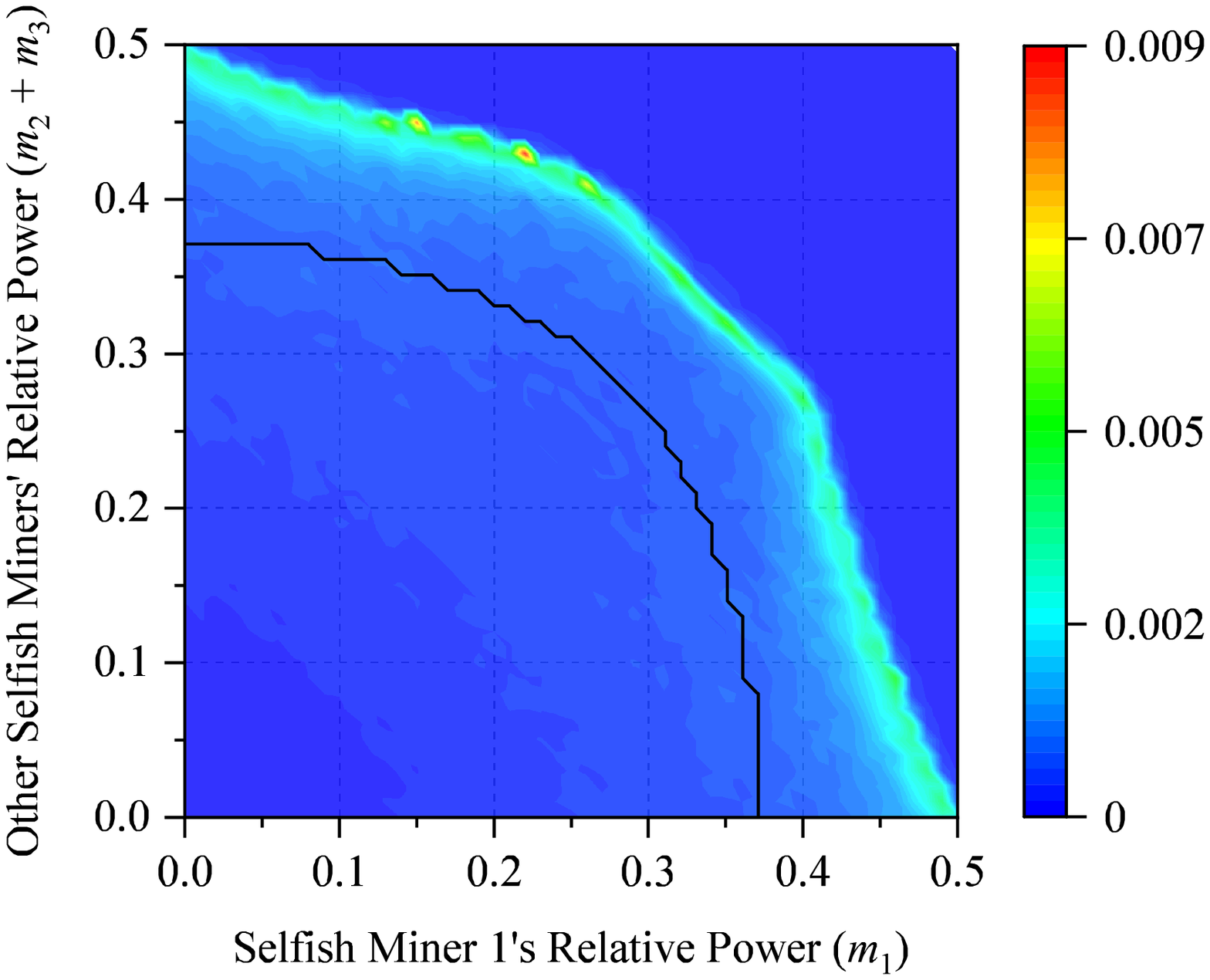}
  \label{fig:3hm_minCI}
 } 
 \subfloat[]{
  \centering
  \includegraphics[width=0.23\textwidth]{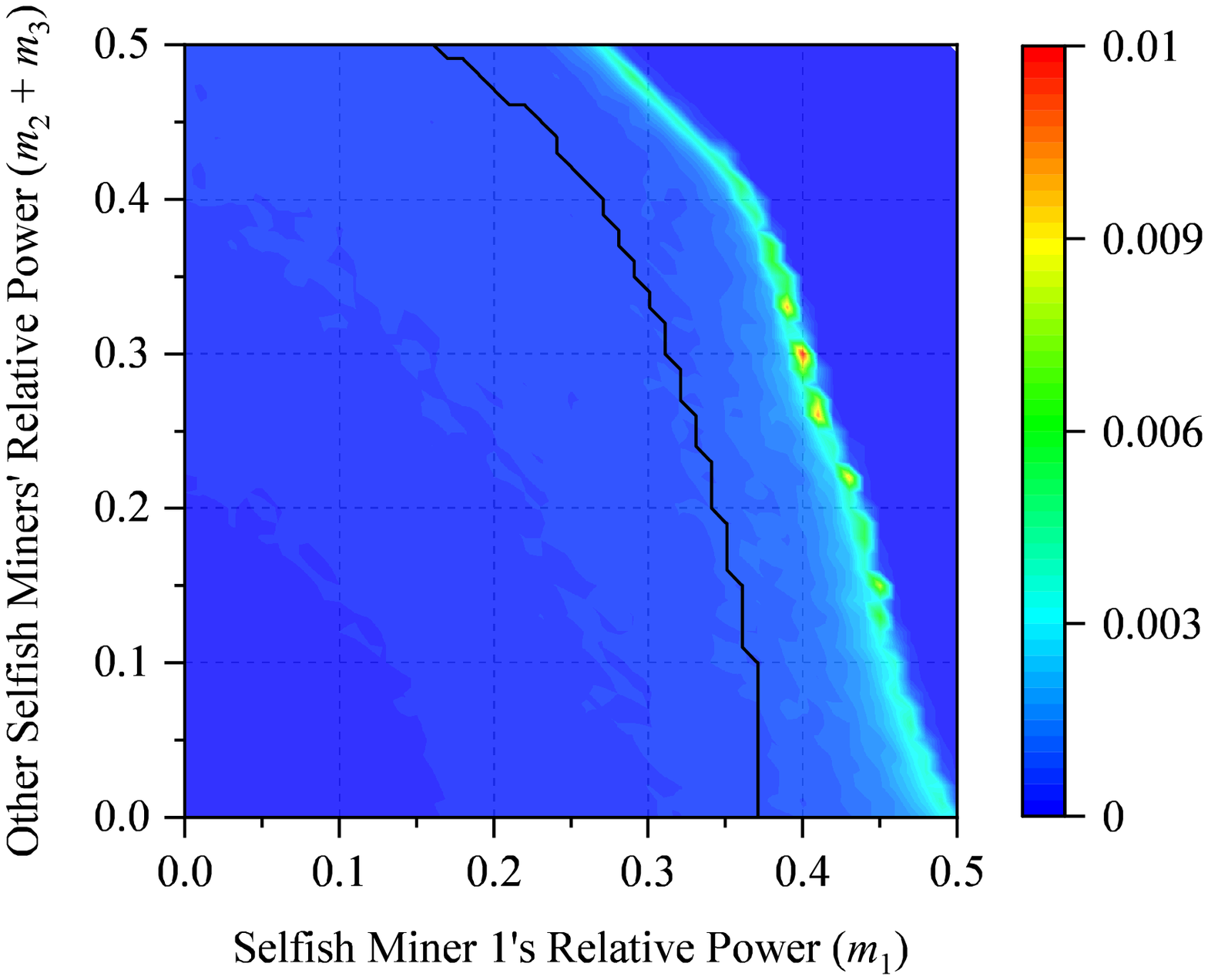}
  \label{fig:3hm_maxCI}
 }
 \caption{Heat maps of selfish miner 1's minimum and maximum average reward (a,b), honest miner's minimum and maximum average reward (c,d), and their 95\% confidence intervals respectively (e-h) in our model with 3 selfish miners. In each plot of selfish miner 1's (a,b,e,f), a black line separates power configurations that yields his relative mining reward greater than his relative power to the right side and vice versa to the left side of the plot. Similarly, a black line in each plot of honest miner's (b,d,g,h) separates power configurations that yields a relative mining reward at least equal to his relative power to the bottom left side and vice versa to the top right side of the plot.}
 \label{fig:3}
\end{figure}

Due to our limitation of graphically presenting the whole results under this setting, only maximum reward and minimum reward of selfish miner 1 are shown in figure \ref{fig:3}, where we vary selfish miner 1's relative power and the other selfish miners' combining power. Particular results of our interest are described in this section.

Unexpectedly, a power threshold that is required for selfish mining strategy becomes even lower in comparison to the previous setting. Specifically, a selfish miner needs a relative mining power of at very least 0.23 to gain his unfair amount of mining reward. Corresponding power configurations are listed in table \ref{tab:3sm} and a visual overview of selfish miner 1's reward is demonstrated in figure \ref{fig:3}. Other configurations $M = \left\lbrace 0.23 , m_2 , m_3 , m_4 \right\rbrace $ where $ m_2 $ and $ m_3 $ are outside of the reported range do not yield extra reward for selfish miner 1. Under such configurations, it is possible that selfish miner 1 does not have enough mining power to build the longest blockchain.

\begin{table}
 \caption{List of power configurations in our model with 3 selfish miners where selfish miner 1 has the least amount of mining power but still earns his unfair amount of mining reward.}
 \label{tab:3sm}
 \centering
 \small
 \begin{tabular}{ccc}
  \toprule
  $ m_1 $ & $ m_2 $ & Range of $ m_3 $ \\
  \midrule
  \multirow{11}{*}{0.23} & 0.15 & $ \left[ 0.21 , 0.23 \right] $ \\
  & 0.16 & $ \left[ 0.18 , 0.24 \right] $ \\
  & 0.17 & $ \left[ 0.17 , 0.24 \right] $ \\
  & 0.18 & $ \left[ 0.16 , 0.25 \right] $ \\
  & 0.19 & $ \left[ 0.16 , 0.25 \right] $ \\
  & 0.20 & $ \left[ 0.16 , 0.25 \right] $ \\
  & 0.21 & $ \left[ 0.15 , 0.25 \right] $ \\
  & 0.22 & $ \left[ 0.15 , 0.24 \right] $ \\
  & 0.23 & $ \left[ 0.15 , 0.24 \right] $ \\
  & 0.24 & $ \left[ 0.16 , 0.23 \right] $ \\
  & 0.25 & $ \left[ 0.18 , 0.21 \right] $ \\
  \bottomrule
 \end{tabular}
\end{table}

Moreover, all selfish miners can simultaneously earn their unfair amounts of mining reward under some particular power configurations. As portrayed in figure \ref{fig:3sm_symm_200rep}, a power configuration $ M = \left\lbrace m_1 , m_2 , m_3 , m_4 \right\rbrace $ in which $ m_1 \in \left[ 0.23, 0.33 \right] $, $ m_1 = m_2 = m_3 $, and $m_4$ has the rest allows all selfish miners to gain their relative rewards greater than their powers. Although, their reward amounts are unsteady when their relative powers are more than or equal to 0.25. Furthermore, their relative rewards could be lower than their relative power once they possess power greater than or equal to 0.26. Note that there are also power configurations where only two selfish miners gain their extra amounts of reward but the another does not.

In addition, we noticed that (i) a system under this setting where an honest miner has relative power lower than 0.34 is always prone to selfish mining regardless of any selfish miner's power configuration, and (ii) there exist some power configurations $ M = \left\lbrace m_1, m_2, m_3, m_4 \right\rbrace $ where $ m_4 \in \left[ 0.32, 0.61 \right] $ such that an honest miner still earns at least his fair amount of mining reward in proportion to his mining power whilst at least one selfish miner gains his relative reward more than his relative power.

Although we do not have a simulation result of the conventional model with 3 selfish miners, it can be expected that a power threshold of selfish mining strategy in our model is slightly greater than one of the conventional model similarly to the previous setting.

\section{Discussion}

\begin{table}
 \caption{A summary of selfish mining strategy's power threshold, a range of selfish miner's equal powers constituting Nash equilibria, and a safety level of honest miner's power to secure Bitcoin system against selfish mining strategy with respect to a number of selfish miners in our model.}
 \label{tab:summary}
 \scriptsize
 \centering
 \begin{tabular}{ccccccc}
  \toprule
  \multirow{3}{*}{\begin{tabular}{@{}c@{}} Number of \\ Selfish Miners \end{tabular}} & \multicolumn{2}{c}{Power Threshold} & \multicolumn{2}{c}{Range of Selfish Miners' Equal Power} & \multicolumn{2}{c}{Safety Level} \\
  \cmidrule(r){2-7}
   & \begin{tabular}{@{}c@{}} Lower \\ Bound \end{tabular} & \begin{tabular}{@{}c@{}} Upper \\ Bound \end{tabular} & \begin{tabular}{@{}c@{}} Stable \\ Nash Equilibria \end{tabular} & \begin{tabular}{@{}c@{}} Unstable \\ Nash Equilibria \end{tabular} & \begin{tabular}{@{}c@{}} Lower \\ Bound \end{tabular} & \begin{tabular}{@{}c@{}} Upper \\ Bound \end{tabular} \\
  \midrule
  1 & 0.38* & 0.38* & - & - & 0.63* & 0.63* \\
  2 & 0.29 & 0.50 & $\left[ 0.29 , 0.40 \right]$ & $\left[ 0.41, 0.49 \right] $ & 0.44 & 0.63 \\
  3 & 0.23 & 0.50 & $\left[ 0.23 , 0.25 \right]$ & $\left[ 0.26, 0.33 \right] $ & 0.34 & 0.63 \\
  \bottomrule
  \multicolumn{7}{l}{
  \multirow{4}{*}{\begin{tabular}{@{}l@{}} 
  * Due to a low granularity of varying power configurations in our simulation, a relative mining \\ 
  \ \ power 0.38 allows a selfish miner to gain his unfair amount of reward, but a power 0.37 does \\
  \ \ not. As a consequence, a safety level becomes 0.63. By a non-linear interpolation, an exact \\
  \ \ power threshold for effectively mounting a selfish mining strategy is approximately 0.3786. \\
  \end{tabular}}} \\
  \\
  \\
  \\
  \\
 \end{tabular}
\end{table}

As summarised in table ~\ref{tab:summary}, a lower bound of power threshold or the least amount of mining power required for selfish mining strategy decreases with respect to an increase of a number of selfish miners in the system. All power configurations that result in the lower bound demonstrate the same trait: a selfish miner has to possess mining power large enough to frequently win against an honest miner and other selfish miners in a race to create the longest blockchain. 


Generally speaking, the more power a selfish miner has beyond the lower bound of power threshold, the higher possibility he will earn his unfair amount of mining reward. That is, the number of power configurations in which a selfish mining strategy is effective increases in proportion to an amount of mining power that the selfish miner has.

However, not every power configuration for a specific amount of selfish miner's power between the lower bound and the upper bound of power threshold always allows him to gain his relative reward greater than his relative power. In other words, a selfish mining strategy might not be effective if there is another miner who possesses mining power to a certain degree greater than the selfish miner himself. As a consequence, a low-power miner using selfish mining strategy is less likely to frequently create the longest blockchain and gain his extra mining reward.

Interestingly, there always exists a power configuration where all selfish miners in the system can simultaneously earn their extra amounts of relative reward and henceforth it constitutes a Nash equilibrium. Particularly, such configuration requires all selfish miners to have an equal and sufficiently high amount of mining power. Since all selfish miners under such configuration would be worse off switching to an honest mining strategy, a Nash equilibrium in which all miners do not change to another strategy is formed. 

Nevertheless, a power configuration where selfish miners have equally yet excessively high mining power might not form a stable Nash equilibria. Under such configuration, all selfish miners have an equal probability of creating a block at each timestep. However, it is entirely  up to chance which selfish miner consecutively get new blocks and consequently build up his blockchain longer than the others'. Furthermore, a selfish miner who could not create the longest blockchain at the previous attempt might be successful in the subsequent attempt. As such, it results in an unsteady amount of mining reward which could be lower than their relative mining power. Therefore, the Nash equilibria that are formed under such configurations are unstable.

From a perspective of security, an amount of mining power to make Bitcoin system completely secure remains the same. As shown in the table \ref{tab:summary}, an upper bound of safety level (or the minimum amount of honest miner's power that prevents any configuration of selfish miners' power) remains constant regardless of any number of selfish miners in the system. An explanation of the cause is intuitive: there is only one power configuration that requires an honest miner to possess a relatively great amount of mining power to prevent selfish mining, namely a configuration where only one selfish miner possesses mining power but other selfish miners do not. Clearly such configuration corresponds to a one-selfish-miner setting (or a case where all selfish miners pool their mining power and work together against an honest miner), and hence results in the same upper bound of safety level for all other settings.

On the contrary, a lower bound of safety level (or the least amount of mining power required to prevent selfish mining in at least one specific power configuration) decreases in proportion to an increase in a number of selfish miners. By further observation, a power configuration corresponding to the lower bound is always one where all selfish miners have power equally but not high enough to gain their unfair amounts of mining reward. It can be implied that they waste their mining power by trying to create the longest blockchain which is eventually replaced by honest miner's.

Finally, it is worth mentioning that our model results in a power threshold slightly greater than one from the conventional model. Though our model of Bitcoin system might not perfectly reflect the real one, it has demonstrated an impact of a concurrency of individual mining processes. As a consequence, all power thresholds of selfish mining strategy and other attacks which were estimated in other works so far could be slightly underestimated.

\section{Conclusion}

In this work, a preliminary investigation of selfish mining strategy employed by multiple miners has been carried out. Based on our empirical results, we have identified that a lower bound of power threshold required to effectively use a selfish mining strategy decreases in proportion to a number of selfish miners in the system. Nevertheless an upper bound of power threshold remains constant, namely 0.5, regardless of a number of selfish miners. 

Another interesting aspect is an existence of Nash equilibria where a number of miners use selfish mining strategy and simultaneously gain their relative mining rewards higher than their relative mining powers. The only required condition for such outcome is an equal amount of every selfish miner's mining power which must be sufficiently large to frequently create a private blockchain longer than honest miner's blockchain. Note that the equilibria could be unstable if their mining powers are excessively high.

On the other hand, a safety level of mining power to be held by non-malicious miners and completely secure Bitcoin system against selfish mining remains the same regardless of a number of selfish miners in Bitcoin system.

Last but not least, whilst our model of Bitcoin system is slightly different to the widely used model, our findings is still valid and demonstrates that a power threshold for selfish mining strategy and other security attacks might be slightly underestimated.

A number of interesting questions remain to be further investigated. As signified in the original study of selfish mining strategy \cite{Eyal2014}, a network capability of selfish miner is also an important factor that affects how much this strategy could be effective. Such aspect will be taken into account in our future work. Moreover, an optimal selfish mining strategy with respect to multiple selfish miners, similar to one in Sapirshtein's \cite{Sapirshtein2017}, is yet not known. With the optimal strategy, it remains to be seen whether our findings is still valid.

\appendix
\section{Simulation Result of Conventional Model in Two-Selfish-Miners Setting} \label{appA} 

\begin{figure*}
 \centering
 \subfloat[]{
  \centering
  \includegraphics[width=0.33\textwidth]{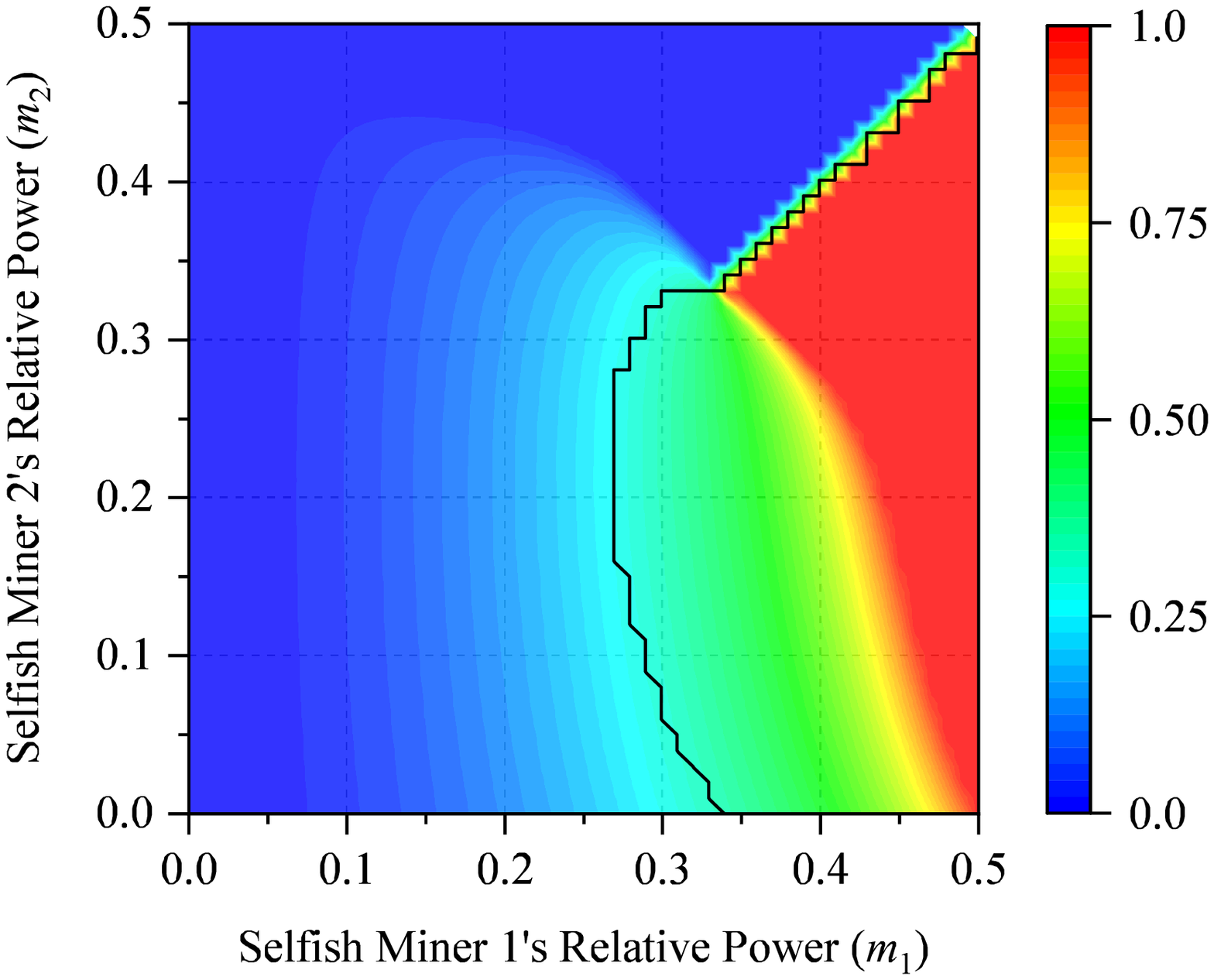}
  \label{fig:2sm_reward_bern}
 } 
 \subfloat[]{
  \centering
  \includegraphics[width=0.33\textwidth]{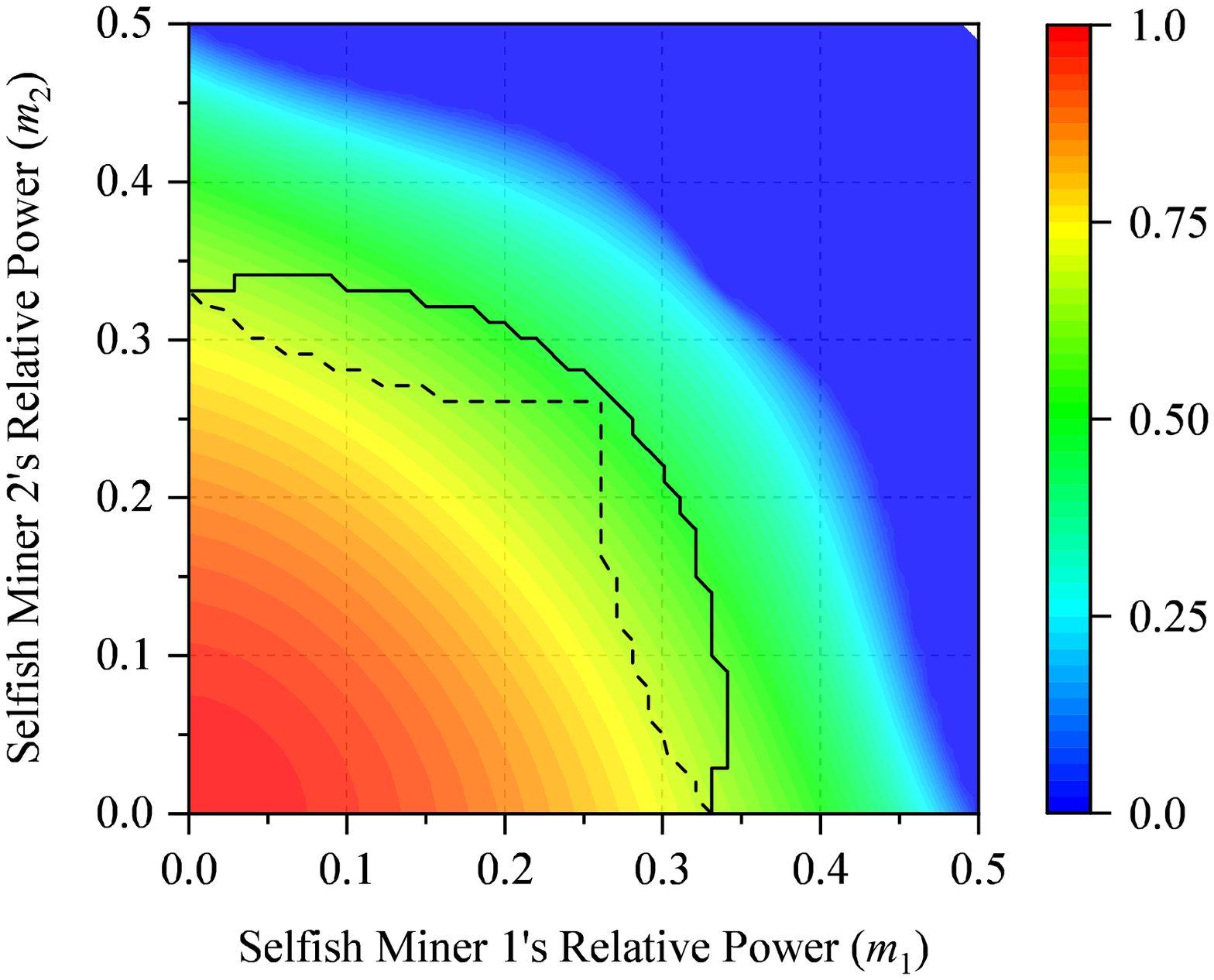}
  \label{fig:2hm_reward_bern}
 } 
 \subfloat[]{
  \centering
  \includegraphics[width=0.33\textwidth]{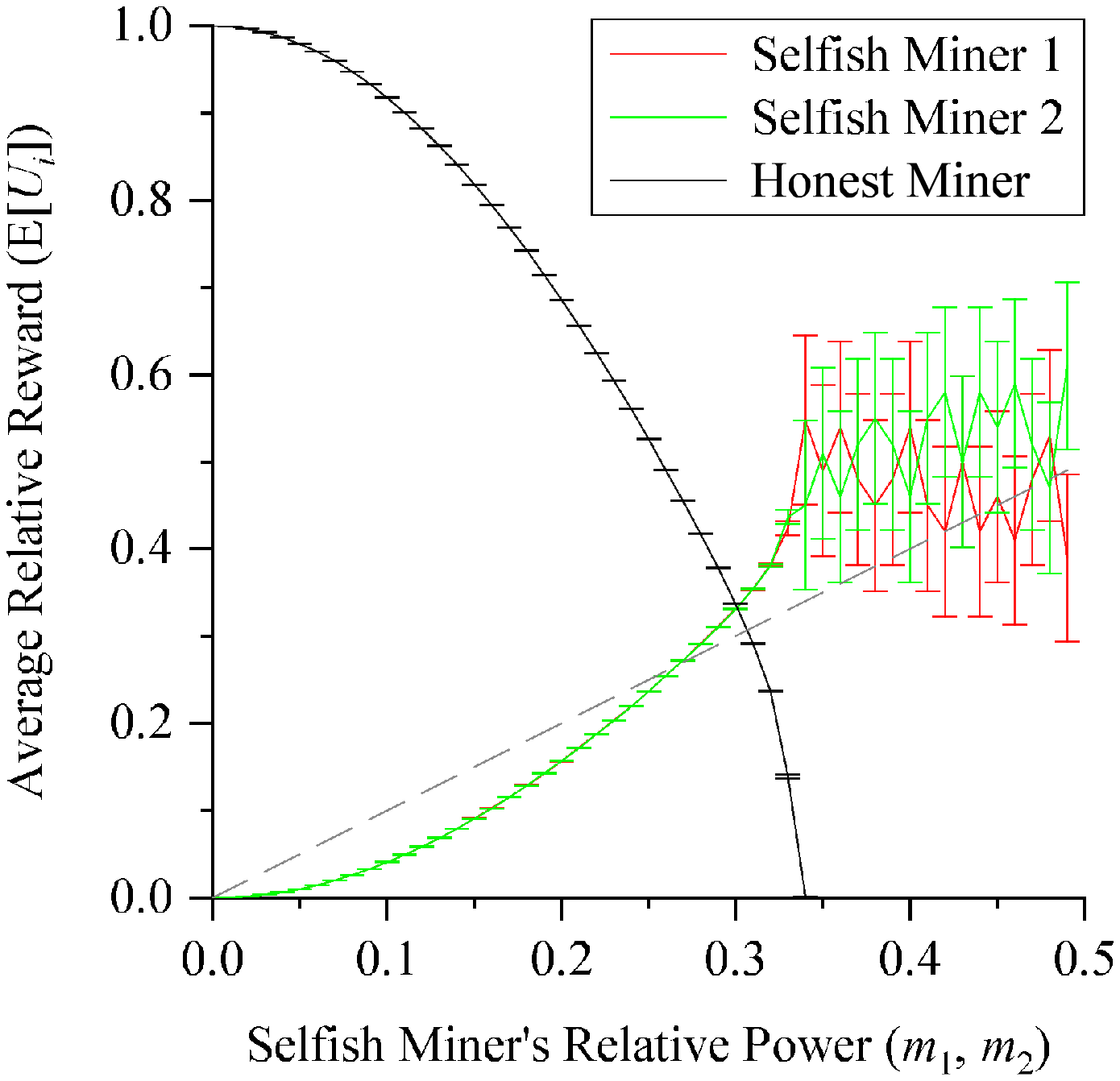}
  \label{fig:2sm_symm_100rep_bern}
 } \\
 \subfloat[]{
  \centering
  \includegraphics[width=0.33\textwidth]{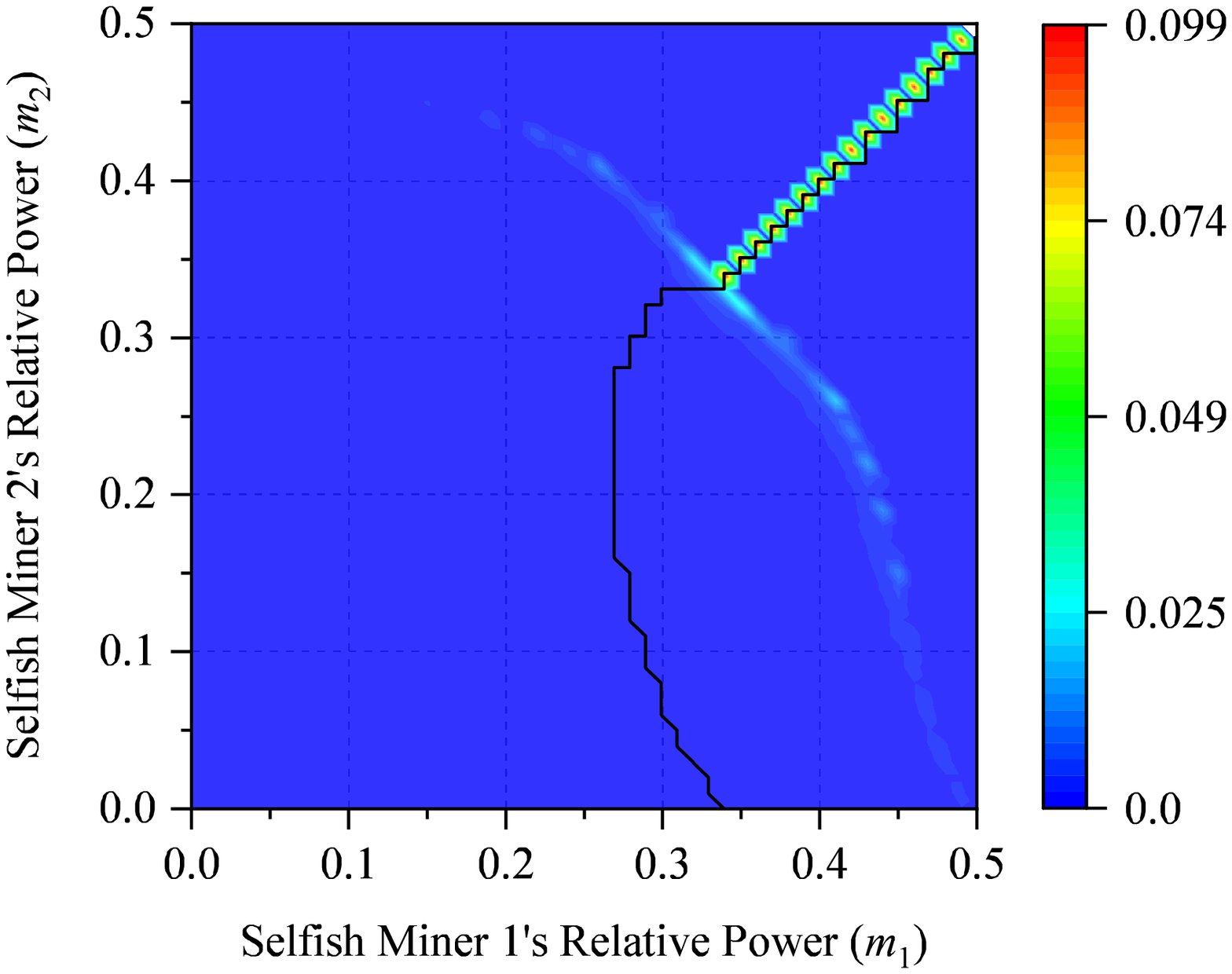}
  \label{fig:2sm_ci_bern}
 } 
 \subfloat[]{
  \centering
  \includegraphics[width=0.33\textwidth]{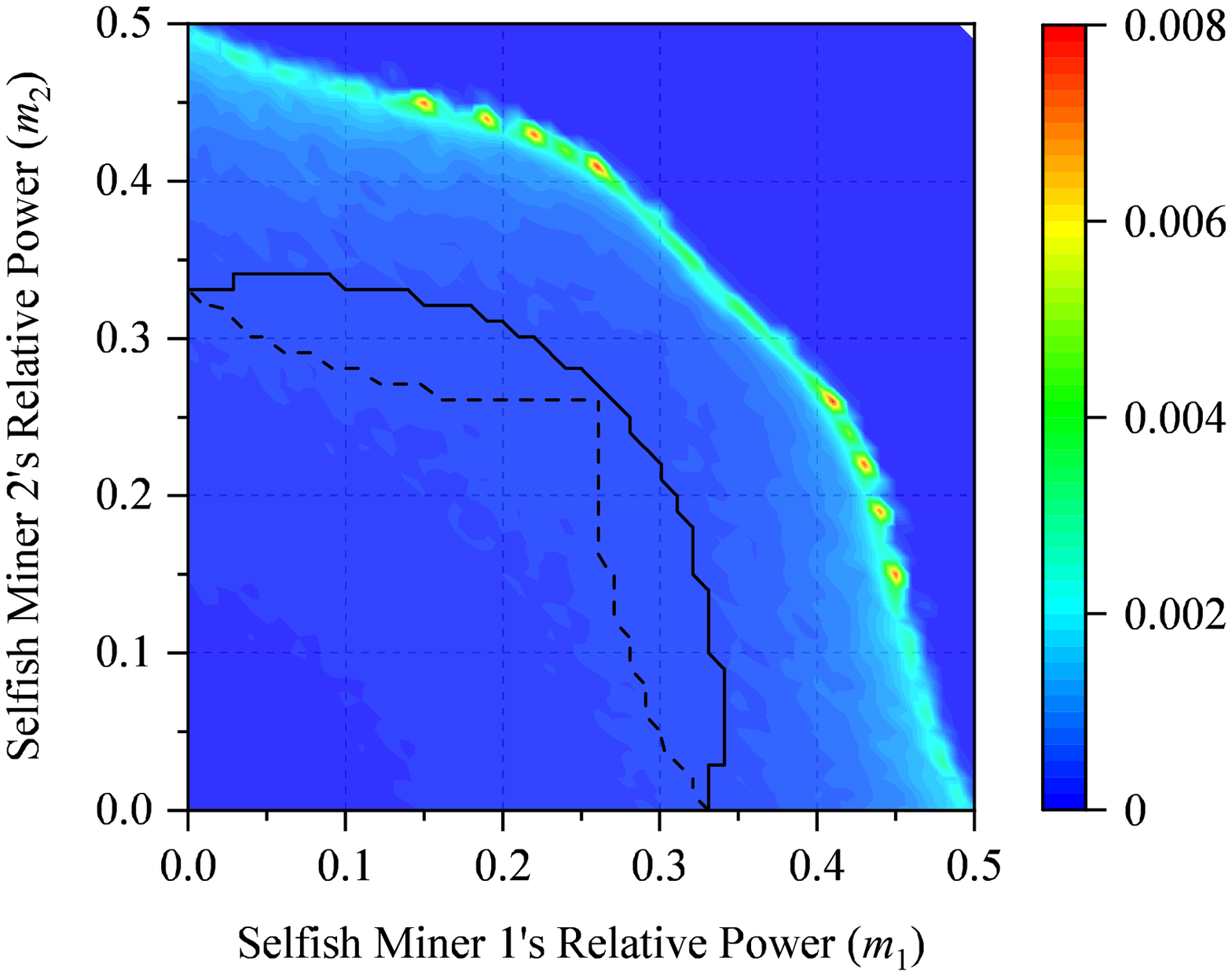}
  \label{fig:2hm_ci_bern}
 }
 \subfloat[]{
  \centering
  \includegraphics[width=0.33\textwidth]{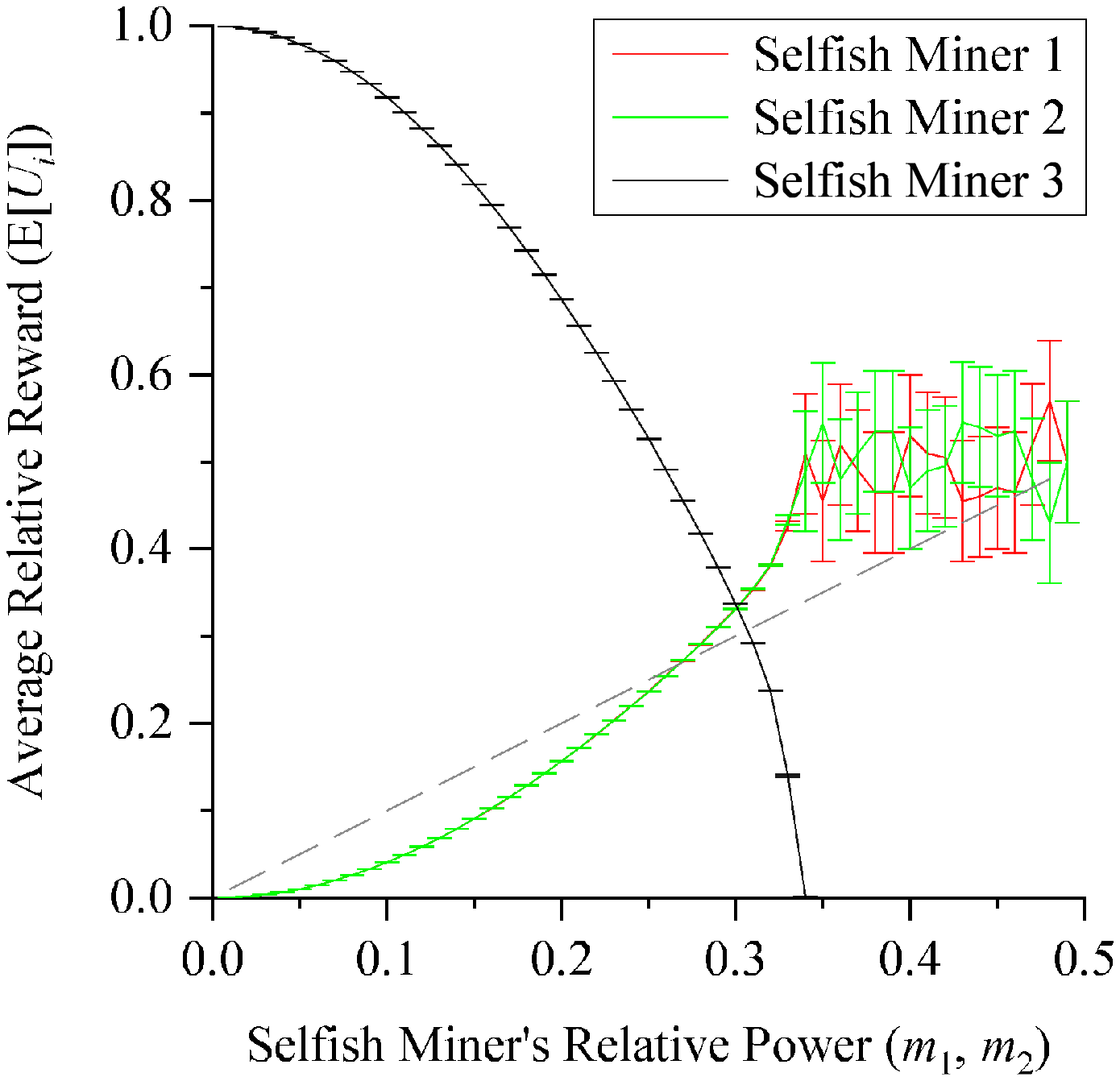}
  \label{fig:2sm_symm_200rep_bern}
 }
 \caption{Heat maps of selfish miner 1's average reward (a), honest miner's average reward (b), their 95\% confidence intervals respectively (d,e); and line plots demonstrating miners' average reward and their 95\% confidence intervals of a power configuration where all selfish miners possess an equal amount of mining power (c,f) in the conventional model with 2 selfish miners. In each plot of selfish miner 1's (a,d), a black line separates power configurations that yield relative mining reward greater than his relative mining power to the right side and vice versa to the left side of the plot. Similarly, a black solid line in each plot of honest miner's (b,e) separates power configurations that yield reward at least equal to his power to the lower left side and vice versa to the upper right side of the plot. Similarly, a dashed line separates configurations that at least 1 selfish miner earns his unfair amount of reward to the upper right side of the plot and vice versa to the lower left side of the plot. As for line plots, plot (c) is a simulation result with 100 repetitions each, whereas plot (f) is a simulation result with 200 repetitions.}
 \label{fig:2_bern}
\end{figure*}

As demonstrated in figure \ref{fig:2_bern}, a power configuration that has the least amount of selfish miner 1's relative power yet allows him to earn his unfair amount of mining reward is $ M = \left\lbrace 0.27, m_2 , m_3 \right\rbrace $ where $ m_2 \in \left[ 0.16 , 0.28 \right] $ and $ m_3 $ has the rest. Other configurations $ M' = \left\lbrace 0.27, m_2 , m_3 \right\rbrace $ where $ m_2 \notin \left[ 0.16 , 0.28 \right] $ do not make selfish miner 1 gain his relative reward higher than his relative power.

In addition, a power configuration $ M = \left\lbrace m_1 , m_2 , m_3 \right\rbrace $ in which $ m_1 \in \left[ 0.27 , 0.49 \right] $, $ m_1 = m_2 $ and $ m_3 $ has the rest allows both selfish miners to simultaneously gain their unfair amounts of mining reward. Similarly to the simulation results of our model, their amounts of mining reward are unstable when both of them equally possess more than 0.33 and could be less than their amounts of relative power if both equally have power greater than or equal to 0.43. Due to a limitation of space in this paper, a plot of these power configurations is omitted.

Finally, a system under this setting is completely defenceless against selfish mining if an honest miner possesses a relative mining power less than or equal to 0.47. In addition, there are some power configurations $ M = \left\lbrace m_1 , m_2 , m_3 \right\rbrace $, $ m_3 \in \left[ 0.47 , 0.54 \right] $ where an honest miner still gains his relative reward no less than his relative power and at least one selfish miner still earns his unfair amount of reward.



\bibliographystyle{plain}
\bibliography{bibliography} 

\end{document}